\documentclass[referee]{aa} 

\usepackage{epsfig}
\usepackage{lscape}

\newcommand{\mic}{\,{\rm \mu m} } 

\begin{document}


\title{FIRBACK IV. Towards the nature of the 170 $\mu$m source population} 

\author{Dennefeld M.\inst{1}\thanks{Visiting Astronomer, Observatoire de Haute-Provence, CNRS, France}, 
Lagache G.\inst{2}, Mei S.\inst{2,3}, Ciliegi P.\inst{4}, Dole H.\inst{2}, 
Mann R.G.\inst{5}, 
Taylor E.L.\inst{5}, Vaccari M.\inst{6}}

\institute{
Institut d'Astrophysique de Paris, 98bis Boulevard Arago, F-75014 Paris \and
Institut d'Astrophysique Spatiale, B\^at.  121, Universit\'e Paris XI, F-91405 Orsay  \and 
Johns Hopkins University, 3400 N. Charles Street, 21218, Baltimore, MD, USA \and
INAF-Osservatorio Astronomico di Bologna, via Ranzani 1, I-40127 Bologna  \and
Institute for Astronomy, University of Edinburg, Royal Observatory , Blackford Hill, Edinburgh, EH9 3HJ United Kingdom \and
Astrophysics Group, Blackett Laboratory, Imperial College, Prince Consort Road, SW7 2AZ London
}

\offprints{dennefel@iap.fr}

\date{Received 13 August 2004; Accepted 14 march 2005}  

\abstract{
We present a detailed study of the brighter ($> 4\sigma$ detections) sources 
in the 170$\mu$m FIRBACK northern N1 ISO survey, with the help of complementary 
data in the 
optical, radio, and mid-IR domain. For 82$\%$ of them,  an optical galaxy  
 counterpart is identified, either as the unique source of the IR emission, 
or as part of a multiple 
identification. With less than 15$\%$ of AGNs, these sources are essentially 
local, moderate starbursters with a dominating cold dust component. They are 
therefore very similar to the galaxies in the IRAS Very Faint Survey or the 
ISO  170$\mu$m Serendipity Survey, and represent a population of cold galaxies 
rather neglected up to now. Their colours do not match those of the  far-IR 
Cosmic IR Background (CIB), to which they contribute  less than 5$\%$. The 
bulk of the sources contributing to the CIB is thus to be searched for in more 
distant galaxies, possibly counterparts of the fainter FIRBACK sources 
still under 
study. These bright, local, galaxies  however play an important role in the 
 evolution of IR galaxies: they dominate the number counts at high  
170 $\mu$m fluxes, and represent  half of the contribution  at 250 mJy. 
Although not particularly massive (typically M$^{*}$), they form more stars 
than a typical spiral galaxy and many are bulge dominated, that
 could represent the remnant of a former merger. 
The fainter part of this population may represent the missing link with the 
higher-z sources found in sub-mm observations. 
}

\authorrunning{M. Dennefeld et al.}
\titlerunning{The  170 $\mu$m FIRBACK sources}

\maketitle
        
\section{Introduction}
The spectral shape of the Cosmic Infrared Background (CIB) detected by
the COBE satellite ({\it FIRAS} experiment between $\sim$150 $\mu$m and $\sim$1 mm,  and 
 {\it DIRBE} experiment at 100, 140 and 
240 $\mu$m)
(\cite{puget96}, see \cite{hauser02}, for a review)
indicates a peak at $\sim150$\,$\mu$m\ with his energy density at least 
comparable to
the optical/UV background.  This peak arises from absorbed optical/UV 
radiation
from star formation and AGN activity in obscured galaxies, 
that is re-radiated in the far-infrared.  This
obscured population of galaxies could host approximately half of the
massive star formation activity over the history of the Universe 
(e.g \cite{flores99}).\\

Together with the Lockman Hole survey (\cite{kawara98}), the FIRBACK 
survey (\cite{dole01}) was the most
reliable and deepest ($\sigma$(170 $\mu$m) $\sim 45$ mJy) infrared census at
wavelengths between 20 $\mu$m\ and  850 $\mu$m,  
and the only far-infrared survey with  a better 
sensitivity  than IRAS, until  {\it SPITZER} observations become available. 
It is  close in wavelength to the peak
of the CIB seen by COBE, and   
because of this wavelength proximity, characterizing the 
FIRBACK sources is an essential step towards understanding the 
nature of the sources that generate the CIB. \\ 

The FIRBACK survey used about 150 hours of ISO observing time, corresponding 
to one of the  largest ISO programs (\cite{kessler00}). It covers about 
4 square degrees 
in three high galactic latitude fields, called FIRBACK South-Marano (FSM), 
FIRBACK North 1 (FN1) 
and FIRBACK North 2 (FN2).  The northern fields observed in FIRBACK 
are a subset of the 
larger area covered by the ELAIS survey at shorter wavelength (\cite{oliv00}). 
The  precise location of the fields, the details of the reduction process, an 
assessment of the reliability of the detected individual sources, 
as well as the positions  of 
the close to  200 sources detected with flux 
S$>$135 mJy (the 3$\sigma$ limit) are given in   \cite{dole01}. 
Our knowledge of these sources is still quite limited. 
 \cite{patris03} have published spectra of the 21 brightest 
FIRBACK sources 
in the southern  FSM field. These bright sources are mostly nearby (z$<$0.2) 
dusty star forming galaxies.
They exhibit star formation rates of a few tens of solar masses per year 
with typical IR luminosities of about 10$^{11}$ L$_{\odot}$. 
The fraction  of Active Galactic Nuclei (AGN) is  low, around 15\% at most. 
In the north, two optical sources ($I=24$ \& 22) were observed in 
FN1 with ESI on Keck, revealing higher redshift objects ( $z=0.5$ and 0.9, 
respectively (\cite{chapman02}). 
These two sources are identified as UltraLuminous IR  galaxies (ULIRG's), 
with merger 
morphologies and relatively cold dust temperatures. 
Finally, based on a  statistical analysis (\cite{sajina03}), the FN1 
sources appear 
to show a bimodal redshift distribution, with normal star-forming galaxies 
at z$\simeq$0 and a  tail of a much more luminous 
galaxy population at z$\sim$0.4-0.9 . \\ 

In this paper, we present the information available in the FN1 field, for 
all FIRBACK sources  
with flux S$>$4$\sigma$ (=185 mJy) and give some insights on the fainter  
flux population. 
The detailed analysis of all the FIRBACK population  
is not complete up to now: the identification process is a long process 
which requires many complementary observations in various wavelength ranges. 
This paper describes the currently available data in this N1 field,
thus allowing    the
 community   to conduct complementary follow-up observations if 
appropriate. Emphasis is put on the brighter (thus presumably closer) sources 
where the identification is secure. Work is continuing on those sources 
with multiple 
or fainter optical counterparts, which probably  represent a population further 
away. \\ 

The paper is organised as follows.  The complementary 
observations available  to date are presented first (Sect. 2). 
Then, we describe the identification process and give the 
results source by source (Sect 3.). In Sect. 4 and 5, we discuss the optical  
and IR properties of our sample. The IR/radio correlation is discussed 
in Sect. 6, and in Sect. 7, we analyse the star formation rates. 
We finally give some global properties for the population 
of sources  with fluxes between  $3 \sigma$ and $4 \sigma$ in Sect. 8 and 
discuss all   the results in the last section (Sect. 9). 
 
\section{Complementary observations} 
The FN1 field has been covered (sometimes in part only) by radio, millimeter,  
sub-millimeter, 90, 15 and 6.7 $\mu$m, observations, U, g', r', i', Z
 and  K-band  imaging and visible spectroscopy. 
In this section, we describe briefly all these data and their 
relevance for the present analysis. 
 
\subsection{The ELAIS observations} 
The European Large Area ISO Survey (ELAIS) has surveyed a total of 12 
square degrees at 15 $\mu$m using the
ISOCAM instrument (Cesarsky et al. 1996) and at 90 $\mu$m using ISOPHOT 
(Lemke et al. 1996). Seven square degrees have  also been covered  at 
6.7 $\mu$m, but only one square degree  at 170 $\mu$m (the FN2 field).
The ELAIS final band-merged catalog has  recently become available 
 (\cite{mrr04}). \\
The FIRBACK-FN1 field discussed here has been covered at 170 $\mu$m over 
two square degrees  in an additionnal  observing run conducted during the 
"supplementary" lifetime of the ISO satellite. 
For  association of  FIRBACK sources in the FN1 field with shorter wavelength
data,  only the ELAIS data at 15 $\mu$m have been used, as it is the only 
wavelength where the field coverage is nearly complete.   
In general, we prefer not to use the 90~$\mu$m 
catalogue (\cite{herau04}) 
since it is clearly incomplete, although, in  some cases, 
positions of
90~$\mu$m sources with high fluxes have been  used to make  
 the identification process converge.
Since the ELAIS survey is a shallow survey, the sensitivity at 15 $\mu$m is 
limited 
by the instrumental noise. 
The catalogue of  \cite{vaca05}  
 (http://astro.imperial.ac.uk/~vaccari/elais),  
which is  used here, has a 
completeness limit at the faint end of about 1 mJy at 15~$\mu$m, with 
a 1$\sigma$ photometric error better than  0.25 mJy. It provides a source 
density of about 170 sources per square degree on average, down to the 
5 $\sigma$ limit, to ease    the FIRBACK identifications. 
Its rather low sensitivity biases 
the 15~$\mu$m detections towards the lowest redshift sources, 
but corresponds well to  the brigthest FIRBACK 
population. 

\subsection{The radio data} 
 The Very Large Array (VLA) has been used by \cite{ciliegi99} 
in C configuration to carry-out a sensitive 20-cm radio 
survey of the FN1, N2/ELAIS and N3/ELAIS fields. 
In the FN1 field, the flux limit varies over the area surveyed:  
a circular central region with a flat noise distribution
($\sim$200 arcmin$^2$, 5$\sigma$=0.135 mJy), surrounded by concentric annular regions,
where the noise increases for increasing distance from the centre
(the last annulus has a 5$\sigma$ rms noise of 1.15 mJy). 
These regions of the sky had also been  surveyed previously 
to shallower flux limits at 20cm, in the NVSS survey
 in the  VLA D configuration (\cite{condon98})  and the FIRST survey 
in the  VLA B configuration (\cite{beck95}). \\ 

For the identification process,  we use both the 5$\sigma$ radio catalogue of 
\cite{ciliegi99} and the FIRST catalogue.
When no radio source is found in the FN1 position error circle, we go back  
to the radio map  to compute 5$\sigma$ upper limits and search for radio sources 
down to the local 3$\sigma$ noise value.  
Within the error circles of  the 56 brightest FIRBACK FN1 sources, 
we found at least one radio source for 30 of them. Eight of those 
 have their radio flux in the  $3<S<5 \sigma$ range, and three have been 
found in the FIRST catalogue.  
When the FN1 source 
position is outside the \cite{ciliegi99} coverage, 
a 5$\sigma$ upper limit from the FIRST survey 
is used.

\subsection{Millimeter and sub-millimeter data} 
 IRAM 1.3 mm observations have been conducted with MAMBO on some ($\sim$10)  radio 
positions that were coincident with  FIRBACK positions. 
Only  2 or 3  sources
seem to be detected. The data are currently under analysis and 
will not be used in this paper.\\ 
 
On the sub-millimeter side, early  observations of some FIRBACK sources were 
published by \cite{scott00}, followed by observations of a sample of 
30 FIRBACK N1 
sources by \cite{sajina03}. This sample as a whole (co-adding all the data 
for all the observed sources) is detected  
at the 10.6$\sigma$ level at 850 $\mu$m and at the 9.0$\sigma$ level at  
450~$\mu$m ($<S_{850}>=2.5 \pm 0.2$~mJy 
and $<S_{450}>=16.7 \pm 1.9$~mJy). Out of the 30 N1 sources, 
7 are detected 
at the $>$3$\sigma$ level at 850~$\mu$m (3 of them have $3 \sigma < S < 4 \sigma$ ) 
and 5 sources are detected 
at the $>$3$\sigma$ level at 450~$\mu$m. 
Only these $S >3 \sigma$ detections will be used here. 
 
\subsection{X-ray data}
A cross-correlation between the ELAIS-ISO 15~$\mu$m survey and the ROSAT 
all-sky survey has been conducted by \cite{basil02}. Three sources were found, 
corresponding roughly to the fields of the 170~$\mu$m sources FN1-35, 126 and 295. 
But the position of the X-ray source is in these 3 cases always at more than 140" from the 
centroid of the PHOT source, and is therefore largely outside our error circle. We can 
therefore conclude that there is no association of a   170~$\mu$m source with an X-ray source,  
at the sensitivity level of the Rosat all-sky survey. \\
Another cross- correlation of the 15~$\mu$m survey 
with specific Chandra pointings in the N1 (and N2) field 
has been performed by \cite{manners04}. Three matches were found in N1, but 
none  correspondings to a FIRBACK source (except perhaps $FN1-042$, see 
the discussion of individual sources). 

\subsection{ Other IR data. }

We derived IRAS 60 and 100 $\mu$m fluxes (or upper limits) for each FN1 
source position,  by using 
 SCANPI{\footnote{http://irsa.ipac.caltech.edu/applications/Scanpi/}}, 
a tool developed for visualizing, plotting and averaging  
calibrated IRAS survey scans.  Each  FN1 source was  checked by eye. 
For point sources with signal-to-noise greater than 3, we determine the 
flux by fitting  the point source template. 
When no clear detection was obtained, a 3$\sigma$ upper limit 
was set. 
The whole procedure is fully described at: 
http://irsa.ipac.caltech.edu/IRASdocs/scanpi/. 
None of the identified FIRBACK galaxies is resolved by IRAS. 
We checked that the fluxes derived  by using SCANPI (in and near 
the FIRBACK fields)  were in good agreement with  
those of the IRAS Faint Source Catalog for sources in common. \\ 
 
We also searched for associations of optical counterparts with 
2MASS (\cite{kleinm94}) sources and an association was found in the 
Extended Source Catalogue (XSC) in most cases. Although several magnitudes 
and colours are usually available, we have extracted only the total 
K$_{s}$ magnitude: we do indeed not intend to use the near-IR colours 
to fit the spectral energy distributions (in view of the probable complex stellar 
populations content), but simply use the K luminosity to estimate the  mass 
of the galaxy. In a few cases when no 2MASS magnitude was available, 
we used K magnitudes from the work  of \cite{sajina03}. \\

The N1 field is one of the fields of the SWIRE legacy survey 
(\cite{lonsd03}) using the SPITZER satellite (\cite{werner04}) 
and is the first one to have been observed. The products (Version 1.0)  
 comprise images and catalogues in the four IRAC bands 
(3.6, 4.5, 5.8 and 8.0 $\mu$m) and three MIPS bands (24, 70 and 160 $\mu$m), 
described by \cite{sura04} on the Spitzer Web site 
(http://ssc.spitzer.caltech.edu/legacy/),
 and  were released on  Oct. 27, 2004, after this paper was first submitted. 
The use of those data for the analysis of FIRBACK sources is thus defered to 
a subsequent paper. We note however that the available 
MIPS 160 $\mu$m catalogue contains 178 entries down to the cut-off limit 
of 200 mJy, in an area of 8.5 square degrees. 
Twenty-five  of them only lie within the smaller  
FN1 area, to be compared to 44 FIRBACK sources down to the same flux 
limit (at 170 instead of 160 $\mu$m). We have however checked on the SWIRE 
160 $\mu$m images that all the FIRBACK sources were present down to 200 mJy, 
and almost all of them (only three possibly  suspicious cases) down 
to 140 mJy, so that all the FIRBACK sources discussed in this paper  are real 
and that the difference 
in number counts  is to be ascribed to 
incompleteness of the preliminary SWIRE 160 $\mu$m N1 catalogue, and 
probably incomplete field coverage.  
While the sensitivities of both surveys are therefore comparable, 
the SWIRE data will later allow a larger sample of presumably similar objects 
to be studied, with the help of the shorter wavelength data for more precise 
identifications. 

\subsection{Optical data} 
Multicolour photometry  was obtained over the FN1 field,  in U through Z, 
as part of the INT Wide Field Survey
{\footnote{http://www.ast.cam.ac.uk/~mike/casu/WFCsur/WFCsur.html}} 
(\cite{mac01}), with  
 the Wide Field Camera covering about 0.3 deg$^2$ at once on the 2.5m Isaac 
Newton Telescope.  
The photometry is described in \cite{gonz05} and the limiting magnitudes 
 range from $\sim$21.9 in Z to $\sim$ 24.9 in g'. The corresponding 
catalogue was used to search for optical associations with the FN1 
sources using the likelihood ratio method, as described most  
recently in \cite{mann02}. 
For each source,   the probability P$_{ran}$ is computed  
that an association with an optical object at the   given 
likelihood ratio has occurred by chance,  on the basis of simulations using 
associations with random positions in the INT data. A low  P$_{ran}$ value 
  therefore means a more secure identification. 
The results of these computations are used together  with   
the 15~$\mu$m and 21 cm catalogs to find the best 
identifications (as explained in Sect \ref{IDS}). 
\\ 
 DSS2 data  were  used for initial identification purposes and preparation 
of the spectroscopic observations, 
waiting the accessibility of the INT images,
and  are shown  as small charts  in Fig. \ref{stamps}.
\\ 
Optical spectra were taken at the Haute-Provence Observatory (CNRS, France) 
with the 1.93m telescope during several observing runs between 1999 and 2004. 
The Carelec long-slit spectrograph was used with a 300 l/mm grating and an 
EEV CCD detector of 2048x1024 pixels of 13.5  $\mu$m each, giving a spectral 
element of 1.75\AA\ per pixel and a spectral resolution of 6.5\AA\ with the 
2$\arcsec$ entrance slit generally used. The total spectral coverage is about 
3600\AA\, but the central wavelength was  different from one run to 
another, to allow  the H$\alpha$ emission sometimes to be detected in 
higher redshift objects. The orientation of the slit was adjusted to 
register more than one object in a given exposure, when adequately bright 
galaxy candidates were available in the PHOT error circle. \\
 The reduction followed the usual procedure, with flat-fielding, wavelength 
calibration, and spectral response determination through observations of 
several standard stars, the master standard  usually being BD+26,2606. The 
relative response is determined with an accuracy of  a few percent, the 
blue part of the spectrum having  the poorest correction due both to the 
decreasing  response of the detector and the  average extinction curve 
available. Absolute fluxes are available only part of the time, due to 
changing weather conditions: when they are, their accuracy is estimated to 
be better than 15$\%$, not taking into account light losses at the entrance 
slit. As these fluxes are only indicative, no correction for those 
losses has been attempted.   \\ 

\section{\label{IDS} Identifications: method and results} 

As the FIRBACK  position error circle (\cite{dole01}) is rather large 
(100 $\arcsec$ in diameter for a 93$\%$ localising probability), the 
identification process is complex. For the bright PHOT sources (most of those 
considered in this paper), we selected the bright galaxy (or galaxies) 
within the PHOT error circle and 
used, as the major criterion, the 
distance between the optical galaxy candidate and the centroid of the PHOT 
detection. If several candidates were possible, the spectral characteristics 
were then used to identify the good ones: the FIRBACK counterparts are 
expected, by analogy with IRAS sources, to be 
primarily starforming galaxies (or AGNs), albeit possibly colder as they are 
selected at longer wavelengths, and their  
 spectral  features should  thus  be rather 
similar to the typical objects found in the IRAS survey (e.g. \cite{vei95} 
and references therein). 
The complementary observations with  
better positional accuracy were used conjointly to find the correct optical 
association(s) to the far-IR source.  We primarily used 
 the ISOCAM 15~$\mu$m and/or the radio 
21~cm data. The ratio of sensitivity between ISOCAM at 
15~$\mu$m (which is instrumental noise limited) and ISOPHOT at 170~$\mu$m 
(which is confusion noise limited)
is such that it is unlikely that 
 an extragalactic  source will be detected at 15~$\mu$m without contributing 
significantly to the FIRBACK 170~$\mu$m flux  
 (unless it has a very unusual SED).  On the other hand, the use 
of the radio data assumes that the detected sources have 
a similar radio to far-IR ratio as  sources previously detected by 
IRAS (e.g. \cite{condon91}). The use of this criterion could prevent 
the detection of a different type of source in this sample (should they 
exist), but in practice, at least for the bright optical galaxies observed 
here, there was in general no ambiguity: the bright galaxy detected closest 
to the PHOT source  indeed had spectral characteristics similar to those of 
IRAS galaxies; it  moreover  very often had an associated 15~$\mic$ detection 
so that the use of the radio data was not essential. It will 
however become important  when going to fainter PHOT sources, where more 
than one, faint, 
optical counterpart can be located within the error circle. \\

\begin{figure*} 
\includegraphics[width=7.5cm,angle=0]{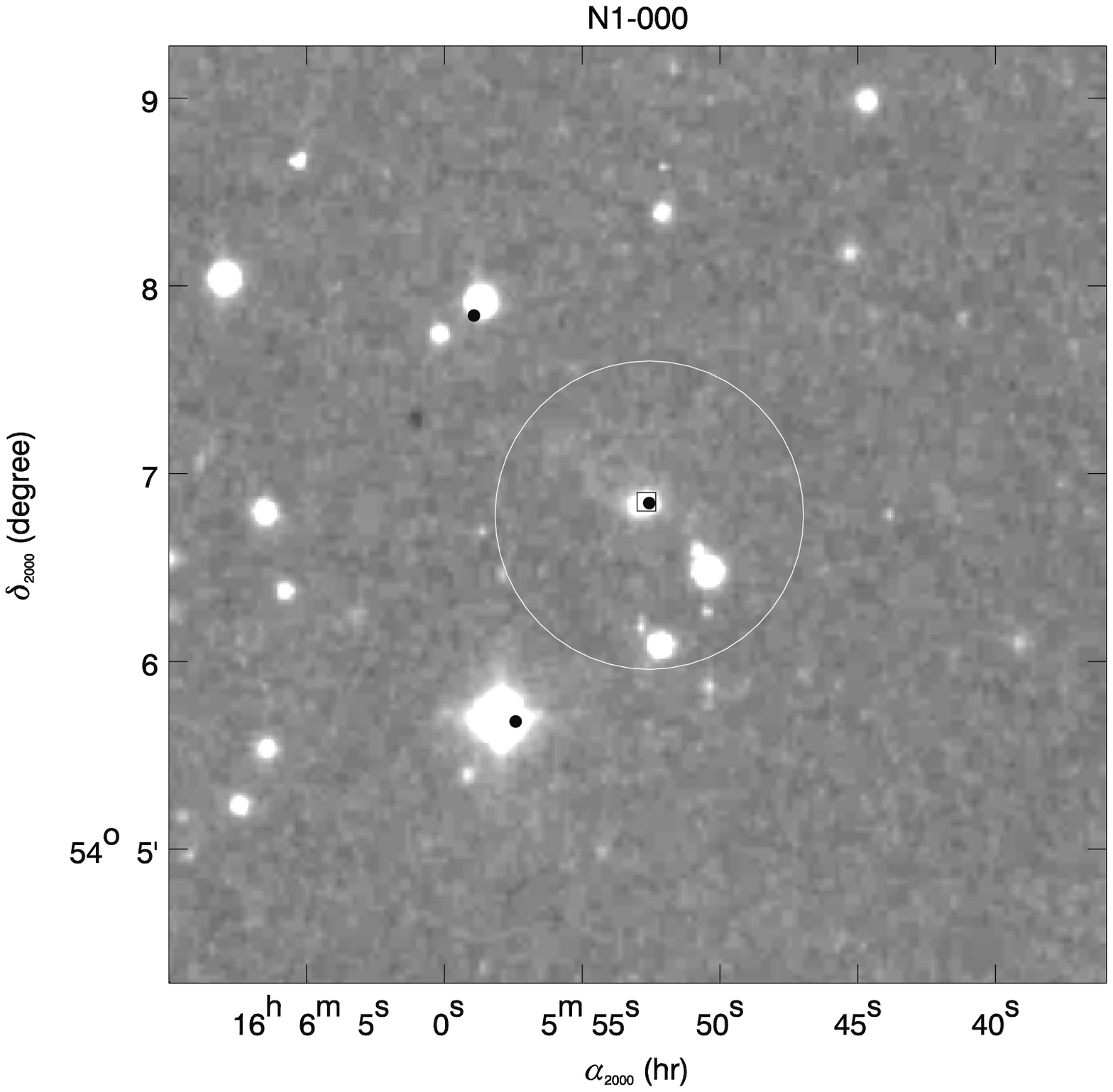}
\includegraphics[width=7.5cm,angle=0]{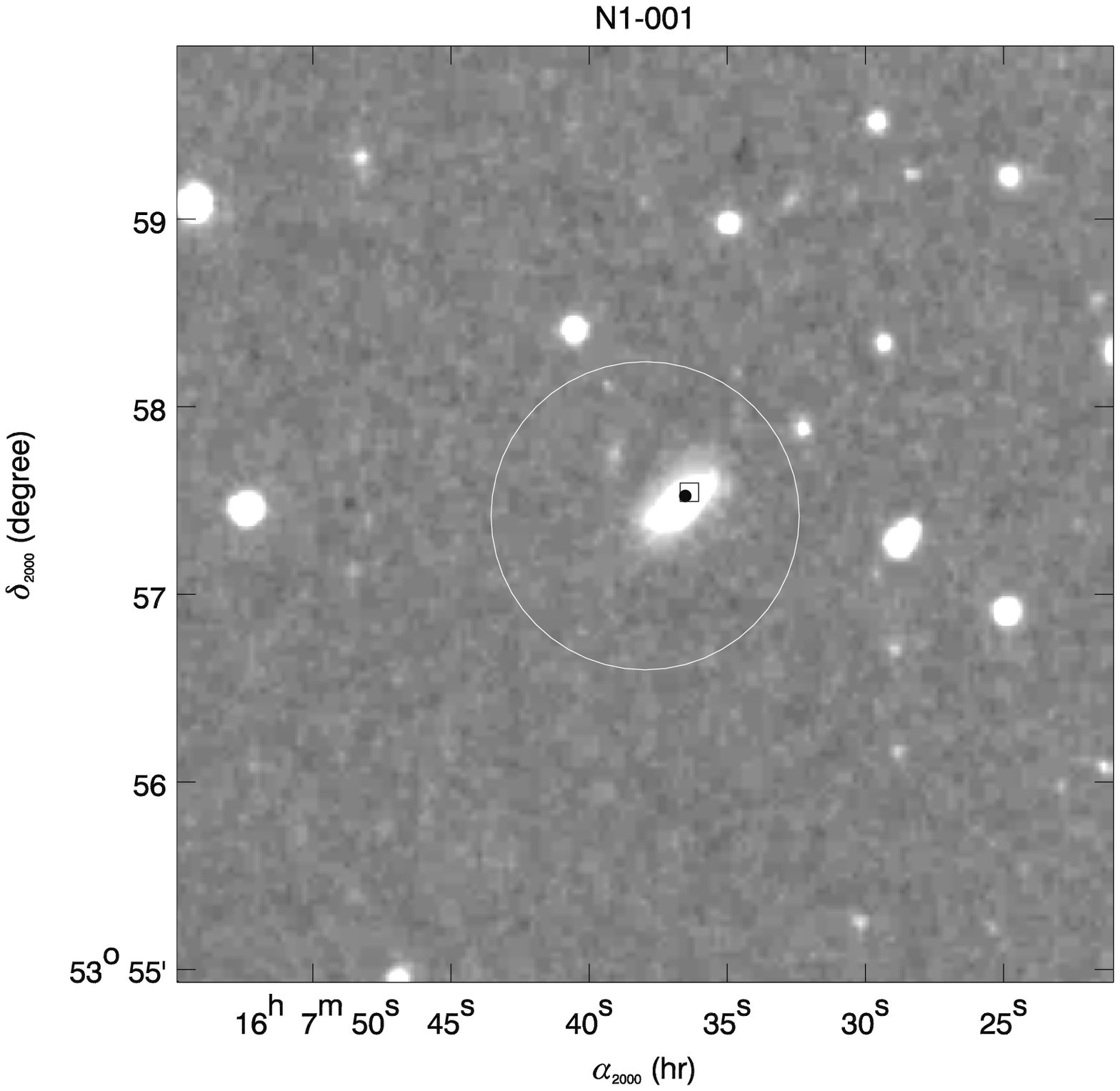}
\caption{Identification charts of  FIRBACK  N1 field 
sources (FN1-0 and FN1-1 as examples). 
 Charts of further objects are only available in electronic form. 
The error circle for the  170$\mu$m ISO
source is superposed on the DSS optical image. An ISO 15$\mu$m source is 
marked by a black dot and a 
 radio source by a small open black square. \label{stamps} }
\end{figure*}

 FIRBACK sources were then distributed in three groups: 
\begin{enumerate} 
\item Identified sources: sources associated with one  bright optical 
galaxy, usually also detected in  radio  and/or ISOCAM.  
Fainter  optical sources, like  FN1-040 (which lies at z=0.45), are also put 
in this category, when the radio/ISOCAM  identification was also unambiguous.   
\item  Multiple sources or uncomplete identification. Into  
 this category are put FIRBACK sources with more than one 
optical/15~$\mu$m/radio   counterpart   
and FIRBACK sources with no secure identification. 
\item Unidentified sources. Either no radio or 15~$\mu$m source detected to 
help identification, or the optical 
galaxies are too faint to be detected spectroscopically with the available 
instrument. We cannot reach a conclusion in such cases.  
\end{enumerate} 
Out of the 56 ISOPHOT sources in the 4 $\sigma$ sample, 28 are fully 
identified, 17 have multiple or incomplete 
identifications and 11 sources have no identified counterpart. 
We comment  on  a few sources in the $3<S<4 \sigma$ range, 
which were also analysed.   
 Fig. \ref{stamps}  presents the DSS images with the FIRBACK error 
circle and the radio and ISOCAM counterparts when present. 

\subsection{\label{identified} Identified sources} 
This section gives details concerning the identified sources. We emphasize 
the complementary data (near-IR or radio) when they are important to secure 
the identification. For all sources, 
fluxes in the various bands (IR and radio) and  far-infrared colors are given 
in Table \ref{main_tab}.
Optical properties are given in Table \ref{opt_tab} and 
spectral characteristics are given in Table \ref{second_tab}. \\ 
$FN1-000$: Identification with a bright optical galaxy, also 
detected at  15 and 90 $\mu$m.   This galaxy is a merger, with clear tidal 
tails. The two other bright 
objects inside the PHOT error circle  are stars. \\ 
$FN1-001$:  Bright optical galaxy. \\
$FN1-002$:  Bright optical galaxy,   also detected at 90~$\mu$m. \\
$FN1-003$:  Bright optical galaxy with 15~$\mu$m emission. 
This identification has a higher  P$_{ran}$ probability to occur by chance, 
because the galaxy is located  $\approx$30~$\arcsec$ from the ISOPHOT 
centroid but  there are no other obvious identifications on the image. 
This source is  detected by IRAS at 100 $\mu$m only (upper 
limit  at 60~$\mu$m, although this band is more sensitive) and should 
therefore be a rather cold galaxy.  
The spectrum shows a low equivalent width of  H${\alpha}$, 
suggesting a weak starburst in an older galaxy. \\
$FN1-004$: Bright optical galaxy with a point-like nucleus. 
This source is also detected  by SCUBA  at 450 $\mu$m 
(32.5$\pm$7.1 mJy). Its spectrum reveals a Seyfert 2 type (strong, narrow
[NII] lines). 
A fainter galaxy at the southern edge of the error circle 
has a similar redshift and also  H${\alpha}$ emission, but is not detected 
in radio nor at 15~$\mu$m: it is unlikely to contribute much to the 
far-IR flux.  \\
$FN1-005$: Faint optical galaxy with 15~$\mu$m detection.
The identification has a higher  P$_{ran}$ value, 
because the galaxy is located  $\approx$30~$\arcsec$ from the ISOPHOT 
centroid, but the ISOCAM identification is unambiguous.  
This object has been observed in the optical 
with the Palomar200/DoubleSpec instrument (Chapman, private communication). \\
$FN1-006$:  bright optical galaxy  detected at 15~$\mu$m and  90~$\mu$m.  
 Because of its location $\approx$30~$\arcsec$ 
away from the PHOT centroid, this identification would have a higher   
probability of occuring by chance, 
but the spectrum is typical of a reddened starburst. 
The bright object just east of the galaxy  is apparently a star. 
A second 15~$\mu$m source, much fainter (a factor of about 6) than the other 
one, is also located within the error circle, but does not show any 
obvious optical counterpart. 
In view of its faintness, and because it does not correspond 
to  the 90~$\mu$m source, it is unlikely to contribute much to 
the  170~$\mu$m flux. \\
$FN1-007$: Bright optical galaxy with 15~$\mu$m, and 1.4GHz emission. 
This source is detected  by   SCUBA 
both at 850~$\mu$m and at   
450$\mu$m  (23.4$\pm$8.1 mJy for the latter). It has a very cold IRAS colour, 
but does not lie in the few, very faint, well identified cirrus filaments in 
the N1 field, so is unlikely to be contaminated by background.  \\ 
$FN1-009$: Identification with a bright optical galaxy with 15~$\mu$m and 1.4GHz emission. 
The spectrum, although dominated by stellar 
features, shows clearly the presence of an active nucleus (strong [OI], large 
[NII]/H$\alpha$ and [SII]/H$\alpha$ ratios), classified as Sey2. The bright 
object at the southern edge of the PHOT error circle, with a 15~$\mu$m 
detection, is a spectroscopically confirmed cold star. \\ 
$FN1-011$:  bright optical galaxy with a 15~$\mu$m detection. 
The other bright object 
to the SW is a star. \\ 
$FN1-012$:  Bright optical galaxy detected  at 15~$\mu$m and 90~$\mu$m. 
   Although the galaxy is 
 $\approx$30~$\arcsec$ from the ISOPHOT centroid, the identification is
secure, with a typical reddened starburst spectrum. \\ 
$FN1-014$: This source does not have 1.4GHz data  
(it lies outside the VLA surveyed area) but has a clear 15~$\mu$m
counterpart and is also detected at 90~$\mu$m. 
  The optical objects were 
however too faint to be observed spectroscopically with our equipment. 
A photometric redshift has been provided by 
T. Babbedge (private communication). \\ 
$FN1-015$: This source is identified with a bright optical galaxy 
with 1.4GHz emission,  and SCUBA detection. It has one of the highest 
redshifts measured in our sample, so that H$\alpha$ falls outside our prime  
spectral range.  \\ 
$FN1-016$: Identification with a bright optical galaxy detected at 15~$\mu$m, 
and 1.4GHz.  This source is detected by IRAS at 60 and 100 $\mu$m, 
and by SCUBA both at 450 and 850~$\mu$m. Although its position is rather 
offset with respect to the PHOT centroid, the identification seems to be 
fairly secure, in view of the detection in all these wavebands, and its 
typical reddened starburst spectrum. \\ 
$FN1-018$: bright optical galaxy with 15~$\mu$m detection but no radio 
emission. Its spectrum, with a strong [NII]/H$\alpha$ ratio, indicates the 
presence of an AGN. \\ 
$FN1-020$: Identification with a faint optical galaxy also detected 
at 15~$\mu$m.  
This source is detected by IRAS at 100 $\mu$m only, and is thus presumably cold. 
 We could not secure 
an optical spectrum, but a photometric redshift has been provided 
by T. Babbedge (private communication). 
 The  other bright object in the error circle seems to be a star. \\
$FN1-021$: bright optical galaxy also detected at  15~$\mu$m. A second galaxy, 
at the eastern edge (but outside, at 69") of the PHOT error circle, also detected 
at 15~$\mu$m, with  a starburst-type spectrum and a velocity close to the 
bright galaxy one, might also contribute to the detected far-IR flux. \\ 
$FN1-023$:  bright optical galaxy with a 15~$\mu$m detection, but not detected 
by IRAS.  The bright object NE of  the galaxy is starlike.  \\ 
$FN1-024$: Bright optical galaxy with 15~$\mu$m and 1.4GHz emission and a 
starburst spectrum. Although there are other, fainter galaxies in the field  
(but not detected in radio nor in mid-IR), this source is considered as 
the main contributor to the far-IR emission. \\ 
$FN1-026$: Identified with  an optical galaxy having also  15~$\mu$m and  
90~$\mu$m detections, but no radio detection.  
The  other two bright  objects in the error circle are presumably stars. \\ 
$FN1-031$: Identified with an optical galaxy with 90~$\mu$m and 1.4GHz 
detections. This source is also detected by  SCUBA at 850~$\mu$m. \\ 
$FN1-033$: Identified with a faint optical galaxy with 15~$\mu$m emission,  
but no radio detection. \\
$FN1-035$: Optical galaxy with also a 15~$\mu$m detection but no radio 
detection (2MASS detection, galaxy classified as IrS). \\ 
$FN1-038$: The faint optical galaxy inside the error circle displays emission 
lines in its spectrum and is associated with a 15~$\mu$m source. 
A  radio source exists, with no obvious optical counterpart, but lies 
outside the error circle and is therefore not associated with the 
PHOT source. \\ 
$FN1-039$: Identified with a faint optical galaxy detected at 1.4GHz. 
This source is also detected by SCUBA at 450~$\mu$m, 
and has been
spectroscopically observed with the Palomar200/DoubleSpec
instrument (Chapman, private communication). \\ 
$FN1-040$: Identified with a faint  optical galaxy detected at  1.4GHz 
and with  SCUBA at 850~$\mu$m. The bright object in the center of the error 
circle is a star. 
This is one of the 2 higher redshift FIRBACK sources detected by 
Chapman et al. (2002), who note it is an interacting pair.  \\
$FN1-041$: Identified with a bright galaxy in the center, detected 
in radio and in mid-IR, with a typical emission-lines spectrum. \\
$FN1-043$: Identified  with an optical galaxy, also detected at  15~$\mu$m.  
A high [NII]/H$\alpha$ ratio 
indicates an active nucleus, but the S/N of the spectrum is insufficient 
to  distinguish between a liner or a reddened Sey2.  \\ 

\subsection{\label{multiple} Multiple or incomplete identifications} 
We detail in this section the FIRBACK sources with more than one possible 
identification and/or a non-secure identification. 
\\ 
$FN1-008$: Two sources  with 1.4GHz emission within the error circle. 
The brightest optical counterpart is at z=0.26 and has a 350 $\mu$m CSO 
detection. It shows a reddened spectrum with absorption features due to hot 
stars, but H$\alpha$ is unfortunately out of our spectral range.  The faintest 
galaxy (the second radio source)   has a SCUBA 
measurement and a K-band magnitude of 14.2 (Sajina et al. 2003) \\ 
$FN1-010$: This source has two 15~$\mu$m counterparts associated with
bright optical galaxies.
It is also detected by IRAS at 60 $\mu$m. The bright 
galaxy in the error circle has a reddened continuum with strong stellar 
absorption features, and H$\alpha$ and [NII] emission of small equivalent 
width. The [NII]/H$\alpha$ ratio clearly indicates an AGN, but the 
non-detection of the emission features in the blue spectral range does not 
allow us to distinguish between its being
 a Sey2 or a Liner. The fainter galaxy within the error circle 
is detected in radio at 1.4GHz and also at 15 and 90 $\mu$m (while the bright 
galaxy is not), but was to faint to be observed spectroscopically here. 
 The  identification of the PHOT source is therefore still uncertain, and 
the far- IR flux may in fact be due  to contributions of both galaxies. \\  
$FN1-017$: This source has a 15 $\mu$m counterpart associated with a faint 
galaxy located at the edge  of the error circle. Many  other faint galaxies 
are visible in the error circle, but they were all too faint to be 
observed here. \\ 
$FN1-019$: Two possible counterparts: one  bright optical galaxy with 
15 and 90~$\mu$m emission, whose spectrum indicates the possible presence 
of a Liner (the emission lines have also a small equivalent width). 
 And one fainter galaxy with 1.4~GHz emission (3-5$\sigma$ 
detection), but no available spectrum. \\ 
$FN1-025$: Two bright galaxies in the error circle display emission lines in 
their spectrum, but are not  related to each other. One has a 15~$\mu$m 
counterpart. 
A fainter galaxy in the field, this one with 1.4~GHz emission 
(3-5$\sigma$ detection), 
could not be observed spectroscopically. The situation is complex, as all three 
galaxies may contribute to the far-IR flux.  \\
$FN1-029$: This source could be identified with an optical galaxy close 
to the PHOT centroid and both 15~$\mu$m and 1.4~GHz detections. 
Another faint, optical galaxy, also with 1.4~GHz emission, lies however at 
the border (51") of  the error circle (not observed spectroscopically) 
and could contribute also to the far-IR emission. \\ 
$FN1-032$: A bright galaxy with 15~$\mu$m counterpart, 
close to the PHOT centroid, displays weak 
emission lines superposed on a strong, reddened, stellar continuum. The 
[NII]/H${\alpha}$ ratio, close to 1, indicates the presence of an active 
nucleus. The 1.4 GHz emission is however  associated with another,  faint 
optical galaxy, also close to the center of the error circle, but which was 
too faint to be observed spectroscopically here. \\  
$FN1-034$: The bright galaxy close to the PHOT centroid displays emission 
lines in its spectrum, has  15$\mu$m emission but no radio counterpart. 
It is also detected by IRAS at both 60 and 100~$\mu$m. 
The radio 1.4GHz emission in this field corresponds 
to a  very faint optical galaxy (K=19.3), with 
a SCUBA detection at 450~$\mu$m (but not at 850~$\mu$m),
which could not be observed spectroscopically. \\ 
$FN1-037$: faint optical galaxy, with also a 15~$\mu$m detection. Could not 
be detected spectroscopically. Another 15~$\mu$m source, inside the error 
circle, is associated with an even fainter galaxy. \\ 
$FN1-042$:  The  bright optical galaxy close to the center (detected in 
mid-IR but not in radio) shows 
only weak  emission lines in its spectrum and is unlikely to be  the real  
counterpart. Three radio sources are detected in the vicinity, two inside 
the error circle, 
associated with faint galaxies which could not be observed in the optical, 
and correspond probably to the emitters of the far-IR flux. 
The third, south-eastern, radio-source is slightly outside the PHOT 
error-circle and corresponds to one of the X-rays sources found by 
\cite{manners04}: they do classify this object as starburst-galaxy, and 
it could contribute also to the far-IR flux. \\
$FN1-044$: A comparatively bright optical galaxy, detected in radio at 
1.4 GHz, and also at 15~$\mu$m and 90~$\mu$m with ISO and 60~$\mu$m with IRAS, 
is undoubtedly a counterpart to the 170~$\mu$m source. Its spectrum 
indicates the presence of an AGN (liner or Sey2). One other radio source 
is however also present within the error circle, and is  associated with a  
very faint 
optical galaxy, which could contribute also to the far-IR emission. \\ 
$FN1-045$: The bright galaxy, with radio emission (and a 2$\sigma$ SCUBA
 detection) is the probable identification. However, there are other, fainter  
galaxies with unknown properties in the field. \\ 
$FN1-046$: Two bright galaxies are close to the border of the FIRBACK error 
circle. The one inside, with no radio emission but 15 $\mu$m emission,  
is a standard emission line galaxy which would be considered as the far-IR 
counterpart if it would be alone. The second galaxy, just outside 
the error circle,  does however not show emission lines.
There is moreover another 15~$\mu$m source
lying in the error circle. This source is associated with a very faint optical 
galaxy. The identification is therefore not complete here. \\ 
$FN1-048$: A bright galaxy, with emission lines, is situated close to the 
center of the error circle. Two radio sources, however, are also present 
(closer to the boundary), one superposed on a bright star (spectroscopically 
confirmed as late type), the other on  a very faint optical galaxy. This latter 
source is also detected by SCUBA both at 450 and 850$\mu$m, and is likely 
  also to be a  contributor to the far-IR source. \\  
$FN1-049$: A bright optical galaxy detected at 15~$\mu$m and at
1.4~GHz (3-5$\sigma$ detection) is certainly
a contributor to the far-IR source. In the error cicle lies another 15~$\mu$m
source associated with a very faint optical object.\\ 
$FN1-050$: Two 15~$\mu$m sources are detected at the limit of the error circle. 
They are both associated with very faint optical galaxies. \\
$FN1-053$: One 15~$\mu$m source is detected at the border of the error
circle (at 49.7 arcsec). It is associated with a very faint optical
galaxy which could not be observed spectroscopically.

\begin{landscape} 
\begin{table*}
\begin{tabular} {|*{10}{c|}}  \hline 
ID& ISO 15 $\mic$ & IRAS 60 $\mic$ & IRAS 100 $\mic$ &  ISO 170 $\mic$ & SCUBA 850  $\mic$ & VLA 1.4 GHz& 100/60& 170/100  \\ \hline 

FN1-000&27.42$\pm$4.17&   620$\pm$ 28&  820$\pm$ 63& 837.7$\pm$ 89.7& & 4.10$^F$$\pm$ 0.15& 1.32& 1.02 \\ \hline 

FN1-001 &17.45$\pm$2.68&260$\pm$31&410 $\pm$99 & 597$\pm$72.5&6.1 $\pm$1.6  &  0.74 $\pm$0.23& 1.58&1.46\\ \hline 

FN1-002&11.67$\pm$1.87 &160$\pm$25&340$\pm$93& 544.5$\pm$68.8& 4.4$\pm$1.1 &0.64$\pm$0.04& 2&1.6\\ \hline 

FN1-003& 8.94$\pm$1.37 & $<$  100 & 220 $\pm$ 60 & 408.3 $\pm$ 59 &  & $<$0.75 & $>$2.2 &1.86 \\ \hline 

FN1-004&11.38$\pm$1.85 &100$\pm$22&300$\pm$73& 390.8$\pm$57.7& 3.6$\pm$1.4 &0.88$\pm$0.10& 3.0&1.3 \\ \hline 

FN1-005&4.15$\pm$0.65&110$\pm$ 34&300$\pm$ 78& 373.8$\pm$ 56.6&     & $<$1.00$^F$& 2.74&1.25 \\ \hline 

FN1-006&11.6$\pm$1.78&100$\pm$ 29& 300$\pm$82 & 347.6$\pm$ 54.7&     & $<$1.00$^F$&3&1.16 \\ \hline 

FN1-007&5.74$\pm$0.9&70$\pm$ 25&520$\pm$ 90 & 337.6$\pm$ 54&4.4$\pm$ 1.6&1.04$\pm$ 0.10& 7.4&0.65 \\ \hline 

FN1-009 &5.71$\pm$0.88 & $<$90  & $<400$ & 313.1 $\pm$  52.3 & 3.5$\pm$1.5 & 1.15 $\pm$  0.07 & &$>$0.78 \\ \hline 

FN1-011&2.78$\pm$0.47&$<$78&$<$255 & 304.5$\pm$ 51.6&      & $<$1.00$^F$ & &$>$1.19 \\ \hline 

FN1-012&3.93$\pm$0.61&90 $\pm$ 34&$<$400& 302.1$\pm$ 51.4& 1.5$\pm$1.6    & 0.31$\pm$  0.07& $<$4.44   &  $>$0.76\\ \hline 

FN1-014& 2.84$\pm$0.48 & 150$\pm$19 & 380$\pm$57 & 295.2$\pm$50.9 & & $<$1.00$^F$ & 2.53 & 0.78 \\ \hline 

FN1-015&2.9$\pm$0.48&$<$78 & $<$300 & 294.4$\pm$50.9 & 1.4$\pm$1.6 & 0.52$\pm$0.07& $>$3.77 & $>$0.98   \\ \hline 

FN1-016&7.87$\pm$1.20& 160$\pm$ 17&360$\pm$ 92& 289.3$\pm$ 50.5& 1.5$\pm$1.2 &1.55$\pm$ 0.07& 2.25&0.8 \\ \hline 

FN1-018&2.77$\pm$0.46&$<$105&$<$210& 287.8$\pm$ 50.4&     & $<$1.00$^F$ &&$>$1.37 \\ \hline 

FN1-020& 2.05$\pm$0.37& $<$  108.& 230.$\pm$76&   283.40 $\pm$ 50.1    &  &   $<$    0.5& $>$2.62 & 1.23 \\ \hline 

FN1-021& 2.98$\pm$0.35& 130$\pm$26 & $<$360 & 271.3$\pm$  49.2& & $<$ 0.75 &$<$ 2.77& $>$0.75\\ \hline 

FN1-023&3.44$\pm$0.55& $<$  69&   $<$294 &269.9 $\pm$ 49.2   &   &$<$1.00  &&   $>$ 0.92 \\ \hline 

FN1-024&5.02$\pm$0.78& $<$84& $<$147& 266.2$\pm$48.9& 2.3$\pm$1.3 & 0.75$\pm$0.04& & $>$1.81 \\ \hline 

FN1-026&9.16$\pm$0.23& $<$  66  &$<$243& 240.6 $\pm$ 47.1&      & $<$1.00& & $>$0.99 \\ \hline 

FN1-031 &3.74$\pm$0.59  & $<$69  & $<$237  & 225.2 $\pm$ 46 & 1.9$\pm$1.1 & 0.43 $\pm$ 0.04 & &$>$0.95 \\ \hline 

FN1-033& 3.62$\pm$0.43&  $<$ 90.&  $<$ 330 &  224.1 $\pm$45.9 & &     $<$ 0.5&& $>$0.68 \\ \hline 

FN1-035	&4.22 $\pm$0.67 & $<$ 63.&  $<$ 192 &  218.2$\pm$  45.5	 &     &$<$0.75 	&&$>$1.14 \\ \hline 

FN1-038 &2.17 $\pm$0.38 & &  &  206.9$\pm$  44.7         &     &$<$1    & & \\ \hline

FN1-039&      & $<$  57&$<$246 & 205.2$\pm$ 44.5& -0.1$\pm$2.3  &   0.58$\pm$  0.05& &   $>$0.83 \\ \hline 

FN1-040&& $<$ 75.&  $<$ 153&  204.7 $\pm$ 44.5&   5.4$\pm$ 1.1&         0.33 $\pm$ 0.03&&$>$1.34 \\ \hline 

FN1-041& 4.50$\pm$0.72 & $<$69 & $<$243 & 204.2$\pm$44.5 & -0.1$\pm$2.5 & 0.76$\pm$0.04 & & $>$0.84 \\ \hline 

FN1-043& 2.47$\pm$0.4& $<$  78&   $<$225 & 200.6 $\pm$ 44.3&     & $<$0.75 &     &  $>$0.89 \\ \hline 

\end{tabular}\\ 
$^F$ : FIRST radio survey 
\caption{Infrared and radio  data (all in mJy) for the FIRBACK identified sources, 
together with  infrared  colors. 
\label{main_tab} } 
\end{table*} 
\end{landscape}

\subsection{\label{unID} Unidentified sources}
For all these sources (except 4), there is no radio or ISOCAM 
detection 
inside  the PHOT  error circle, and no obvious galaxy counterpart. 
For sources $FN1-022$,  $FN1-027$, $FN1-051$, 
$FN1-054$, $FN1-055$, 
we have no further  information  to give. 
The 6 last  sources are:\\ 
$FN1-013$: There is a mid-IR galaxy (F$_{15}$=1.36$\pm$0.3~mJy) associated 
with
a very faint optical galaxy. We have no further information on this object.\\
$FN1-028$: Two bright optical galaxies, one at the edge (SE) of the error  
circle and one well inside (NW) have been observed spectroscopically.  
They have recession velocities of  19921 and 19816  km/sec respectively, but  
do not display emission lines . There is no radio nor 15 $\mu$m detection 
to help identification, and the two optical galaxies are unlikely 
the sources of the far-IR emission.  \\ 
$FN1-030$: The only bright optical source in the error circle is a 
spectroscopically confirmed star. \\
$FN1-036$: One ISOCAM position corresponds to a  very bright star. Other 
bright objects seen in the error circle could also be stars  (like the other
15~$\mu$m faint source). \\ 
$FN1-047$: Only stars are seen inside the error circle, the brightest one 
being also an ISOCAM 15~$\mu$m source. \\
$FN1-052$: This source has one radio counterpart (with radio detection  
above 3 sigma) but no clear optical identification. \\

\subsection{\label{3sigma} Some sources fainter than the $4\sigma$ limit}
For the fainter ISOPHOT flux galaxies (47 sources, $3\sigma<S<4\sigma$)), 
radio and mid-infrared counterparts may be expected to become rarer. 
However, this is only 
marginally the case. The ISOPHOT sources without any radio and mid-IR
counterparts represent 23\% of the $S>4\sigma$ sample
and 34\% of the $3\sigma<S<4\sigma$ sample.
Among the 47 ISOPHOT sources 
with fluxes between 3 and 4$\sigma$, 12 have a 15~$\mu$m counterpart, 12
have a radio counterpart and 7 have both. This tends to show that
the catalog is reliable down to the $3\sigma$ level. 
 Two examples  were observed optically: \\
$FN1-057$: A bright galaxy inside the error circle is also detected at 
15~$\mu$m. It displays emission lines and is the likely identification.  
The source is also strong at both 60 and 100~$\mu$m (IRAS detection). \\
$FN1-101$: Bright galaxy with emission lines, close to the center of the 
error circle, also detected at 15 and 90~$\mu$m. It also coincides with 
the only radio source in this field, and is therefore the obvious 
identification of this much fainter 170~$\mu$m source.\\

\begin{figure} 
\begin{center} 
\epsfxsize=9.cm 
\epsfysize=7.cm 
\epsfbox{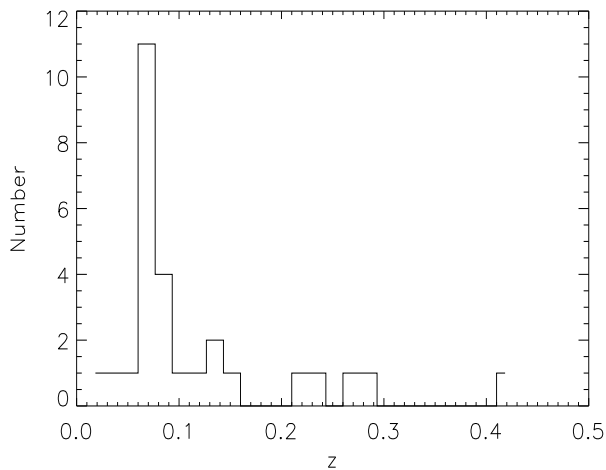} 
\caption{\label{zdis} Redshift distribution of the 28 FIRBACK 
fully identified sources. } 
\end{center} 
\end{figure} 

In all the following (except when specifically mentioned), 
we discusse only sources with secure identifications (28 sources). 

\section{Optical properties of identified sources} 
 
\subsection{Optical colors and morphologies from the INT-WFC survey} 

FIRBACK sources have been observed as part of the INT Wide Field Survey  in 
the U, g', r', i' and Z filters. 
INT photometric bands are similar to the SDSS bands (Sloan Digital Sky Survey 
(Fukugita et al. 1996), see the WFS section  for further details (Sect. 2.6). 

\begin{figure} 
\begin{center} 
\epsfxsize=9.cm 
\epsfbox{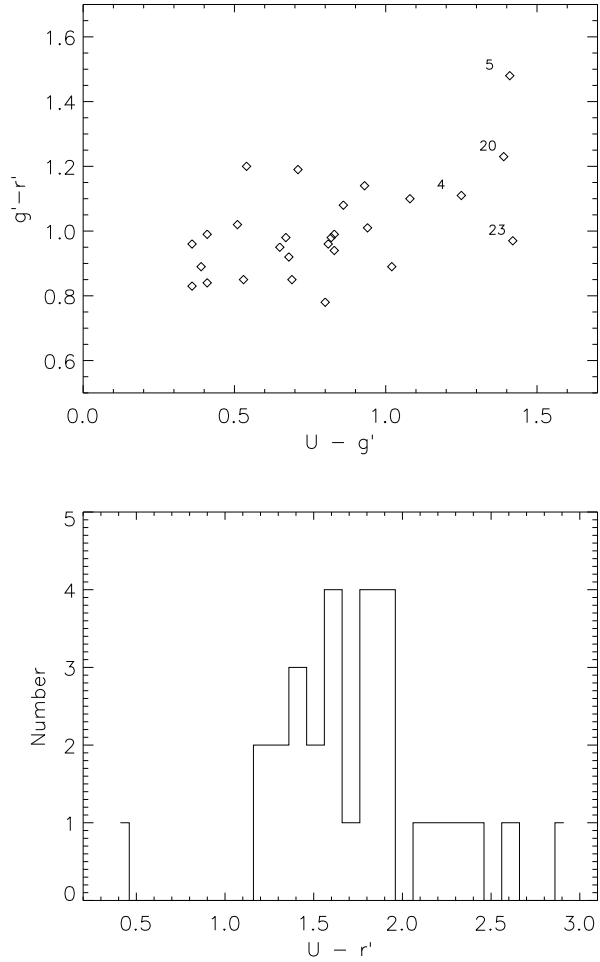} 
\caption{\label{optcol} Optical colors of the identified sources. At the top,  
we show the (g' - r')-(U - r') colour-colour diagram, while the bottom gives 
the (U - r')  histogram.} 
\end{center}
\end{figure}

The morphology of all the objects in our sample was assessed  visually 
on the i' images, using the common Hubble sequence as in Postman et al. 
(2005). 
With the Advanced Camera for Surveys (ACS) on the Hubble Space Telescope, a 
visual classification of that kind has a typical random error of 25$\%$, which 
can be reduced to 6$\%$ when only two broad categories are used (early-type, 
and spirals/irregulars). Although we used ground-based images here, the objects 
are all relatively close so that the uncertainty in classification is of the 
same order of magnitude. \\  
Optical colors were derived for  identified sources and they are given in 
Table~\ref{opt_tab}. 
The source $FN1-040$ is too faint for precise photometry and $FN1-000$ does
not have  a 
 $g'$ magnitude. They are both excluded from  the following analysis. 
The  $( g'-r')$ and the $(U - r')$  color 
histograms are shown in Fig.~\ref{optcol}.  \cite{str01} have found, from 
the analysis of about 150000 galaxies in the SDSS, that 
the distribution of galaxies in  $(U - r')$  is strongly 
bimodal with an optimal color separator of $( U - r') = 2.22$ for low 
redshift ($z < 0.4$) objects:  late  type   
 galaxies lie mostly on the blue side, with  $(U - r') < 2.22$.   
Although selected from the far-IR, most of our galaxies lie also 
on the blue side: 
this is coherent with their classification as late-type for most of them, but  
suggests that they are not heavily obscured, at least outside the central 
regions.  
The reddest sources ($( U - r') > 2.2$; 
FN1-04, FN1-05, FN1-20, FN1-23) are also spiral galaxies,  
but this time probably affected by heavier extinction. 

These measurements have not been corrected for galactic extinction,  but 
this correction should be  negligeable
 in such low HI-column density field as the N1 field 
(N$_{\rm HI}$= 10$^{20}$ at/cm$^2$).\\
Uncertainties on the magnitudes are smaller than 0.1 mag. \\ 

\begin{landscape} 
\begin{table*}
\begin{tabular} {|*{13}{c|}}  \hline 

ID&  Position (J2000)  &  P$_{ran}$ & r' Mag. & (U-r') & (U-g')& (g'-r')&  K$_{s}$ mag & Morphology & Comments\\ \hline

FN1-000 & 160552.8$+$540651 &0.07 &17.94 &     &     & 1.46 & 13.41 $\pm$0.13  & Irregular   & Merger with tail \\ \hline 

FN1-001 & 160736.6$+$535730 &0.00 &16.67 &1.28 &0.39 & 0.89 & 12.29 $\pm$0.09  &  Sb/c       & \\ \hline

FN1-002 & 161005.8$+$541029 &0.10 &17.73 &1.6  &0.68 & 0.92 & 12.63 $\pm$0.11  &  Sb/c       & \\ \hline

FN1-003 & 161254.1$+$545526 &0.29 &17.13 &1.94 &0.86 & 1.08 & 12.57 $\pm$0.09  &  Sa         & Em. with small EW \\ \hline

FN1-004N& 161109.5$+$535808 &0.01 &16.84 & 2.36&1.25 & 1.11 & 11.80 $\pm$0.07  &  Sa/b       & Star forming knots (Seyfert) \\ 
FN1-004S& 161107.3$+$535711 & 1   &19.03 &1.59 &0.74 & 0.85 &                  &  Sa/b       &            \\ \hline

FN1-005 & 160443.4$+$543331 & 0.84& 19.22& 2.89&1.41 & 1.48 &                  & 2 Spirals   & Merger \\ \hline

FN1-006 & 160433.9$+$544432 &0.32 &17.26 &1.77 &0.83 & 0.94 & 12.79 $\pm$0.10  & Sa/O        &\\ \hline

FN1-007 & 161331.2$+$541630 &0.03 &17.18 &1.91 &1.02 & 0.89 & 12.64 $\pm$0.09  & SO/a        & \\ \hline

FN1-009 & 160803.7$+$545302 & 0.01& 16.58& 1.58&0.80 & 0.78 & 11.88 $\pm$0.07  & Sa          & Liner \\ \hline

FN1-011 & 160808.6$+$535242 & 0.13&17.84 & 1.77&0.81 & 0.96 & 13.21 $\pm$0.16  & Sa          &\\ \hline

FN1-012 & 161214.0$+$540833 & 0.30&17.88 &1.38 &0.53 & 0.85 & 13.69 $\pm$0.18  & Sb          &\\ \hline

FN1-014 & 161551.4$+$541536 & 0.43& 19.20&1.32 &0.36 & 0.96 &                  & Sa/b        &\\ \hline

FN1-015 & 160724.7$+$541212 & 0.37&18.89 &1.53 &0.51 & 1.02 & 14.8 $\pm$0.1 $^a$   & Sa/b        &\\ \hline 

FN1-016 & 16-738.0$+$544602 &0.97&17.09  & 1.80&0.82 & 0.98 & 13.14 $\pm$0.10                & Elliptical  & \ \\ \hline

FN1-018 & 161407.1$+$541920 &0.87&18.92  & 2.18&1.08 & 1.1  & 13.31 $\pm$0.20  & Sa/b        & Edge-on (AGN)  \\ \hline

FN1-020 & 160812.3$+$545526 &0.88&19.72  &2.62 &1.39 & 1.23 &                  & Disk?       & Very faint \\ \hline

FN1-021 & 161308.0$+$545142 &0.33&18.18  &1.60 &0.65 & 0.95 & 13.05 $\pm$0.12  & Sa/b        & \\ 
FN1-021B& 161319.2$+$545137 &1   & 18.04 & 0.44&-0.10& 0.54 &                  & E/SO        & \\ \hline

\end{tabular}\\
$^a$: K magnitude from  \cite{sajina03}
\caption{\label{opt_tab} Photometric  properties and morphology of the 
FIRBACK fully identified  sources.  }
\end{table*} 
\end{landscape}
\addtocounter{table}{-1}

\begin{landscape} 
\begin{table*}
\begin{tabular} {|*{13}{c|}}  \hline 
ID& Position (J2000)  &  P$_{ran}$ & r' Mag. & (U-r') & (U-g')& (g'-r')&  K$_{s}$ mag & Morphology & Comments\\ \hline

FN1-023& 160835.4+535020 &0.40 &18.94 &2.39 &1.42 &0.97 & 13.63 $\pm$0.19      & Sb/c   &\\ \hline

FN1-024& 160937.5+541259 &0.54 &18.36 &2.07 &0.93 &1.14 & 13.50 $\pm$0.15      & SO   &\\ \hline 

FN1-026& 161436.9+541635 &0.27 &19.10 &1.82 &0.83 &0.99 & 13.75 $\pm$0.18      & Sb     & Edge-on, dust lanes \\ \hline

FN1-031& 161103.7+544322 &0.05 &17.69 &1.25 &0.41 &0.84 & 13.36 $\pm$0.16      & SO/a ? & \\ \hline

FN1-033& 161301.1+541003 &0.22 &18.47 &1.40 &0.41 &0.99 &                      & E/SO   & \\ \hline

FN1-035& 161527.7+543414 &0.39 &18.11 &1.65 &0.67 &0.98 & 13.38 $\pm$0.14      & Sa/b   & \\ \hline

FN1-038& 160747.3+534836 &0.53 &19.26 &1.85 &0.93 &0.92 &                      & Sa/b   & \\ \hline
 
FN1-039& 160848.9+545151 &0.59 &20.18 &1.90 &0.71 &1.19 & 15.8 $\pm$0.2 $^a$   & ?      & Possibly disk \\ \hline

FN1-041& 160814.2+542836 &0.30 &18.89 &1.19 &0.36 &0.83 & 14.7 $\pm$0.1 $^a$   & SO/a   & \\ \hline

FN1-043& 160553.4+542226 &0.59 &18.69 &1.74 &0.54 &1.20 & 14.00 $\pm$0.20      & Sb     & Barred \\ \hline

\end{tabular}\\
$^a$: K magnitude from  \cite{sajina03}
\caption{Photometric  properties (continued)... }
\end{table*} 
\end{landscape}

\begin{landscape}
\begin{table*}[p]
\begin{tabular} {|*{13}{c|}}  \hline 

ID&  Position (J2000)  &  P$_{ran}$ & r' Mag. & (U-r') & (U-g')& (g'-r')&  K$_{s}$ mag& Morphology & Comments\\ \hline

FN1-008 & 160858.0$+$541818 &0.21 &18.83 & 2.25 & 0.81 & 1.44  & 13.76 $\pm$0.19  & SO/a ?  & Two objects  \\ \hline

FN1-010 & 160930.9$+$535148 &0.26 &17.18 & 2.09 & 1.04 & 1.05  & 12.63 $\pm$0.14  & Sa/b    & Prob. AGN  \\ \hline

FN1-019 & 161235.4$+$541545 &0.01 &17.05 & 1.95 & 0.94 & 1.01  & 12.31 $\pm$0.10  & Sa/b    & AGN  \\ \hline

FN1-025N& 160833.4$+$545510 &0.88 &18.81 & 1.43 & 0.64 & 0.79  &                  & Sb/c    & \\  
FN1-025S& 160834.5$+$545421 &??   &      &      &      &       &                  & Sb      & \\ \hline

FN1-028A& 160738.2$+$534250 &0.47 &17.59 & 2.04 & 1.05 & 0.99  & 13.68 $\pm$0.18  & SO/a    & \\  
FN1-028B& 160746.5$+$534151 &1    &16.70 & 2.20 & 1.18 & 1.02  & 12.26 $\pm$0.09  & SO      & \\ \hline 

FN1-029 & 161117.5$+$541629 &0.26 &18.67 & 2.38 & 1.38 & 1.00  & 13.65 $\pm$0.18  & Sa/b    &\\ \hline

FN1-032 & 161239.9$+$543657 &0.07 &17.38 & 2.33 & 1.34 & 0.99  & 12.07 $\pm$0.08  & Sa      & AGN  \\ \hline

FN1-034 & 160724.2$+$544330 &0.33 &18.66 & 0.90 & 0.19 & 0.71  & 19.3 $\pm$0.8$^a$& Sa/b    &  \\ \hline

FN1-037 & 161509.0$+$541837 &0.36 &19.42 & 1.88 & 1.14 & 0.74  &                  & Sb      & Id.  uncertain \\ \hline


FN1-042 & 161038.9$+$543628 &0.08 &17.24 & 2.24 & 1.35 & 0.89  & 12.06 $\pm$0.08  & Sb      & AGN  \\ \hline 

FN1-044 & 160931.6$+$541827 &0.74 &18.69 & 1.99 & 0.93 & 1.06  &                  & Sa      & \\ \hline 

FN1-045 & 160853.9$+$544735 &0.6  &19.24 & 2.12 & 1.11 & 1.01  & 14.33 $\pm$0.17  & Sa      & \\ \hline

FN1-046A& 161249.4$+$540837 &1    &18.01 & 2.44 & 1.22 & 1.22  & 13.19 $\pm$0.13  & SO/a    & Radio source \\ 
FN1-046B& 161251.1$+$540801 &0.52 &17.38 & 1.61 & 0.68 & 0.93  & 13.09 $\pm$0.12  & SBa     & Barred  \\ \hline 

FN1-048 & 161059.1$+$542305 &0.64 &19.37 & 1.31 & 0.44 & 0.87  & 19.3 $\pm$0.9$^a$& Sb/c    &  \\ \hline

FN1-049 & 160801.6$+$543642 &0.54 &18.89 & 1.54 & 0.69 & 0.85  & 13.73 $\pm$0.20  &         & \\ \hline

FN1-057 & 160801.6$+$543643 &0.78 &18.43 & 1.64 & 0.69 & 0.95  & 13.26 $\pm$0.15  & SBa     & Barred \\ \hline

FN1-101 & 160946.1$+$542127 &0.11 &18.44 & 0.85 & 0.25 & 0.60  &                  & Sa/b    &\\ \hline
\end{tabular}\\
$^a$: K magnitude from  \cite{sajina03}
\caption{\label{Phot_tab} Photometry  of additional sources 
(sources with multiple or uncomplete identification); or sources outside 
the 4$\sigma$ sample (FN1-057 and FN1-101). }
\end{table*} 
\end{landscape}

\subsection{ Spectral properties} 

Most of the spectra display both emission and absorption lines. All lines
have been used to derive the redshift (except those with low signal to noise 
ratio, indicated in brackets in Table \ref{second_tab} and 
in Table \ref{spec_tab}). 
The 1$\sigma$ dispersion  
 is given after the velocity: it should be  smaller when more lines are 
available, but the well-known systematic difference between emission and 
absorption lines also affects the result when both types of lines are used. To 
this internal error has to be added  the (random) external error, which should 
not exceed one pixel after correction for flexures, so that the final 
uncertainty in velocities should be  better than 150 km/s in most cases.  
The distribution in redshifts of the  identified sample is plotted in 
Fig.~\ref{zdis}. Out of the 28 identified sources, about 80\% have
redshifts lower than 0.25. This is in very good agreement with
the redshift distribution of the \cite{lagache03} model of
evolution of IR galaxies. 
\\
The emission line spectra are very similar to those of standard IRAS 
galaxies (e.g. \cite{vei95}), with the classical Balmer lines 
(H$\alpha$ and H$\beta$ essentially) and forbidden lines of 
oxygen (3727, 5007, and,  
less often, 6300$\AA$), nitrogen (6548, 6584$\AA$) and 
sulfur (6717, 6731$\AA$).  
In most cases, the excitation is low, and the continuum appears to be 
reddened, so that quite often the H$\beta$ or [OIII] lines are not even 
detected. Because the H$\beta$ emission, when detected, is often superposed 
on the corresponding absorption line, the correct estimate of its 
intensity would require a proper determination (and subtraction) of the 
underlying absorption, which was not possible with the available  spectra. 
We therefore have not derived an estimate of the reddening from the 
H$\alpha$/H$\beta$ ratio. This prevents us from deriving reliable  
H$\alpha$ luminosities, even for those objects observed under photometric 
conditions: the values given are therefore only lower limits.  \\
We have not detected any broad Balmer line objects in this sample. The 
signal to noise ratio in the continuum would however  not have allowed us 
to detect faint, broad wings in most cases (for Seyfert types 1.9 or 1.8). 
But we have a number of objects 
where the [NII]/H$\alpha$ ratio is  larger than the usual HII region  
limit of 0.5 
(e.g. \cite{vei02}), and so is the [SII]/H$\alpha$ ratio,  
therefore indicating the presence of an AGN. 
Only exceptionnaly  are other line ratios available, 
preventing  a more precise classification of this AGN: the usual weakness or 
absence of the [OIII] line  however points preferentially towards a Liner 
rather than a Seyfert 2 galaxy. We have 8 such cases 
out of a total of 50 objects, that is 15 $\%$, a proportion similar to the  
one found in other IR-selected samples of low redshift objects with moderate 
IR luminosities. 

\begin{landscape} 
\begin{table*}
\begin{tabular} {|*{10}{c|}}  \hline 

ID& Coordinates$^a$ & $cz$ (km/s)   & r' Mag. &  Emission (H$\alpha$ flux)& Absorption& Comments\\ \hline

FN1-000 & 160552.8$+$540651 &43604 $\pm$87 &17.94 & H$\alpha$(183), [NII],  [OI], [NI], H$\beta$, [OIII] & Na& Liner \\ 
        &                   &              &      & H$\gamma$, [OII], (HeI)                              &   &  \\ \hline

FN1-001 & 160736.6$+$535730 &9004 $\pm$51  &16.67 & H$\alpha$(93.9), [NII], [SII], H$\beta$, [OIII], HeI& Na, Ca, (Mg), G, H, K& \\ \hline

FN1-002 & 161005.8$+$541029 &19119 $\pm$119&17.73 & H$\alpha$(13.9), [NII], [SII], H$\beta$, [OII]& Na, Mg, G, H, K& \\ \hline

FN1-003 & 161254.1$+$545526 &19481 $\pm$90 &17.13 & H$\alpha$(21.7), [NII], [SII], [OII]& Na, (H$\beta$, Mg, Ca, G, H, K& Em. with small EW \\ \hline

FN1-004N& 161109.5$+$535807 &19107 $\pm$85 &16.84 & H$\alpha$(9.5), [NII], [SII], [OIII]& Na, H$\beta$, G, Mg& Sey2, [NII]/H$\alpha$$>$1\\ 
FN1-004S& 161107.2$+$535711 &19208 $\pm$88 &19.03 & H$\alpha$(5.5), [NII], [SII]& Na, G&\\ \hline

FN1-005 & 160443.4$+$543331 &86400$^b$     & 19.22&                             &&\\ \hline

FN1-006 & 160433.9$+$544432 &22608 $\pm$63 &17.26 & H$\alpha$(34.8), [NII],[SII], [OII]& Na, Mg, H$\beta$, G, H$\delta$, H, K&\\ \hline

FN1-007 & 161331.2$+$541630 &18435 $\pm$41 &17.18 & H$\alpha$(59.1), [NII], [SII], [OII]& Na, G, H, K& \\ \hline

FN1-009 & 160803.7$+$545302 &15746 $\pm$865& 16.58& H$\alpha$(17.0), [NII], [SII] & Na, Mg, Ca, H$\beta$, H$\gamma$, H$\delta$ & Liner \\ 
        &                   &              &      &   [OI], [OIII], [OII]& G, H, K&  \\ \hline 

FN1-011 & 160808.6$+$535242 &19326 $\pm$48 &17.84 & H$\alpha$, [NII], [SII], H$\beta$& Na, Ca, G, (H), K&\\ \hline

FN1-012 & 161214.0$+$540833 &19770 $\pm$73 &17.88 & H$\alpha$(32.7), [NII], [SII], H$\beta$, [OIII], [OII]& Na, Mg, Ca, H, K&\\ \hline

FN1-014 & 161551.4$+$541536 &21000$^c$     & 19.20&                   & &\\ \hline

FN1-015 & 160724.7$+$541212 &70183 $\pm$58 &18.89 & H$\beta$, [OII]& H, K&\\ \hline 

FN1-016 & 16-738.0$+$544602 &27579 $\pm$85 &17.09  & H$\alpha$(48.7), [NII], [OII]& Na, (H$\beta$), (H$\gamma$), H$\delta$, H, K& \ \\ \hline

FN1-018 & 161407.1$+$541920 &25598 $\pm$105&18.92  & H$\alpha$(7.9), [NII], ([OIII]), [OII]& Na& AGN ([NII]/H$\alpha$$\sim$1) \\ \hline

FN1-020 & 160812.3$+$545526 &3000$^c$      &19.72  &                 &&\\ \hline

FN1-021 & 161308.0$+$545142 &19349 $\pm$39 &18.18  & H$\alpha$, [NII], [SII], H$\beta$& Na& \\ 
FN1-021B& 161319.2$+$545138 &19983 $\pm$18 & 18.04 & H$\alpha$, [NII], [SII], H$\beta$, [OIII]& & \\ \hline

\end{tabular} \\
$^a$: J2000 position of the optical counterpart.
$^b$: Chapman (private communication)
$^c$: photometric estimate (\cite{babb04})
\caption{\label{second_tab} Spectroscopic properties  of the FIRBACK 
identified sources.  
The quoted uncertainty for velocities is the internal (1$\sigma$) error, 
and does not include the systematics (see text). 
Emission (or absorption) lines shown in brackets are seen, but not used 
for the velocity determination because of poor S/N. 
The H$\alpha$ flux is given, when conditions were photometric, in units of 
$10^{-16} ergs/cm^{2}/s$ .}
\end{table*} 
\addtocounter{table}{-1}
\end{landscape}

\begin{landscape} 
\begin{table*}
\begin{tabular} {|*{10}{c|}}  \hline 
ID& Coordinates & $cz$ (km/s)  & r' Mag. &  Emission (H$\alpha$ flux)& Absorption& Comments\\ \hline

FN1-023& 160835.4+535020 &18780 $\pm$26 &18.94 & H$\alpha$(26.1), [NII], [SII], H$\beta$, [OIII], [OII]& Na&\\ \hline

FN1-024& 160937.5+541259 &25770 $\pm$60 &18.36 & H$\alpha$(20.1), [NII], H$\beta$, ([OII])& Na&\\ \hline 

FN1-026& 161436.9+541635 &23835 $\pm$12 &19.10 & H$\alpha$(4.8), [NII]& (Na), (H$\delta$)&\\ \hline

FN1-031& 161103.7+544322 &18840 $\pm$78 &17.69 & H$\alpha$(8.6), [NII], [SII], ([OIII])& H, K& \\ \hline

FN1-033& 161301.1+541003 &41173 $\pm$65 &18.47 & H$\beta$, [OII]& Na, G, H, (K)& \\ \hline

FN1-035& 161527.7+543414 &40820 $\pm$72 &18.11 & H$\alpha$, [NII], [OII]& Na, Mg, (H$\beta$), H, K&\\ \hline

FN1-038& 160747.3+534836 &32192 $\pm$20 &19.26 & H$\alpha$(10.1), [NII], [SII], H$\beta$, [OIII]& & \\ \hline

FN1-039& 160848.9+545151 &80700$^b$     &20.18 &                              & & \\ \hline

FN1-041& 160814.2+542836 &35995 $\pm$46 &18.89 & H$\alpha$, [NII], H$\beta$, [OIII], H$\gamma$& & \\ \hline

FN1-043& 160553.4+542226 &63470 $\pm$53 &18.69 & H$\alpha$, [NII], [OII]& (G), H&\\ \hline

\end{tabular}\\
\caption{Spectroscopic properties (continued)... For  source FN1-040, 
$z = 0.45$, see Chapman et al. (2002) }
\end{table*} 
\end{landscape}

\begin{landscape}
\begin{table*}[p]
\begin{tabular} {|*{10}{c|}}  \hline 

ID& Coordinates $^a$& $cz$ (km/s)  & r' Mag. &  Emission (H$\alpha$ flux)& Absorption& Comments\\ \hline

FN1-008 & 160858.0$+$541818  &78770 $\pm$83  &18.83   &  [OII]& Na, H, K& \\ \hline

FN1-010 & 160930.9$+$535148  &18741 $\pm$75  &17.18   & H$\alpha$(9.1), [NII], [SII]& Na, Mg, Ca, H$\beta$, G, H, K& Prob. AGN (NII/H$\alpha$$>$0.5) \\ \hline

FN1-019 & 161235.4$+$541545  &18663 $\pm$104 &17.05   & H$\alpha$(5.0), [NII], [OI], [OII]& Na, (Mg), H$\beta$, G, H, K& AGN ([NII]/H$\alpha$$>$0.5 \\ \hline

FN1-025N& 160833.4$+$545510  &15761 $\pm$37  &18.81   & H$\alpha$, [NII], [SII], H$\beta$, [OIII], ([OII])& H& \\  
FN1-025S& 160834.5$+$545421  &32671 $\pm$54  &        & H$\alpha$, [NII], H$\beta$, [OIII], [OII]& & \\ \hline

FN1-028A& 160738.2$+$534250  &19816 $\pm$154 & 17.59  & & Na, Mg, H$\beta$, G, H, K& \\  
FN1-028B& 160746.5$+$534151  &19921 $\pm$142 &16.70   & [OII]& Mg, Ca, (H$\beta$), G, H, K& \\ \hline 

FN1-029 & 161117.5$+$541629  &43100 $\pm$70  &18.67   & H$\alpha$(11.4), [NII]& Na, H, K&\\ \hline

FN1-032 & 161239.9$+$543657  &18674 $\pm$63  &17.38   & H$\alpha$(8.9), [NII], [SII]& Na, Mg, (H$\beta$), G, H, K& AGN (NII/H$\alpha$$\sim$1) \\ \hline

FN1-034 &(160724.2$+$544330) &27916 $\pm$54  &18.66   &   H$\alpha$(27.5), [NII], H$\beta$, [OIII], [OII]& K&  \\ \hline

FN1-037 & 161509.0$+$541837  &85500 $\pm$300 &19.42   & [OII], [NeV]& Mg, K& Id.  uncertain \\ \hline


FN1-042 & 161038.9$+$543628  &18844 $\pm$68  &17.24   & [NII]& Na, H$\beta$, Ca, Mg, H, K& AGN ([NII]/H$\alpha$$>$1) \\ \hline 

FN1-044 & 160931.6$+$541827  &24682 $\pm$41  &18.69   & H$\alpha$(7.2), [NII], [SII], [OIII]& (Na), (G)& \\ \hline 

FN1-045 & 160853.9$+$544735  &276400?        &19.24   & [OII] ??& & Single emission line \\ \hline

FN1-046A& 161249.4$+$540837  &45290 $\pm$16  &18.01   & ([NeV]??)& Na, Mg, (G), H, K& Radio source \\ 
FN1-046B& 161251.1$+$540801  &18566 $\pm$73  &17.38   & H$\alpha$, [NII], [SII], [OII]& Na, G, H, K& \\ \hline 

FN1-048 & 161059.1$+$542305  &18688 $\pm$22  &19.37   & H$\alpha$(10.8), [NII], [SII], H$\beta$, [OIII]& Na&  \\ \hline

FN1-049 & 160801.6$+$543642  &38381 $\pm$107 &18.89   & H$\beta$, [OIII], [OII]& (H), K& \\ \hline

FN1-057 & 160801.6$+$543643  &27546 $\pm$35  &18.43   & H$\alpha$(15.1), [NII]& Na, Ca, (G), H, K& \\ \hline

FN1-101 & 160946.1$+$542127  &19690 $\pm$58  &18.44   & H$\alpha$, [NII], [SII], H$\beta$, [OIII], [OII]& H, K&\\ \hline
\end{tabular}\\
$^a$: J2000 position of the optical counterpart. 
\caption{\label{spec_tab} Spectroscopy of additional sources 
(sources with multiple or uncomplete identification; or sources outside 
the 4$\sigma$ sample (FN1-057 and FN1-101)). The H$\alpha$ 
flux is given in parenthesis, in units of $10^{-16} ergs/cm^{2}/s$, when 
conditions were photometric. }
\end{table*} 
\end{landscape}

\section{Mid and far-IR properties} 

\subsection{For sources with IRAS detections: 60, 100 and 170 colors} 

\begin{figure} 
\begin{center} 
\epsfxsize=9.cm 
\epsfysize=7.cm 
\epsfbox{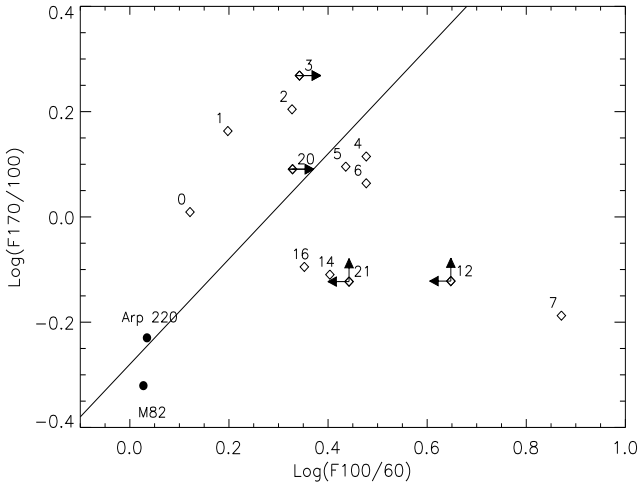} 
\caption{\label{seren} 60, 100, 170 color ratios. The continuous line is the 
ISS fit, see text. } 
\end{center} 
\end{figure} 
The mean infrared (60, 100 and 170~$\mu m$) colors have been calculated for 
 our identified galaxies.  
The two-color diagram log($F_{170\mu m}/F_{100\mu m}$) versus 
log($F_{100\mu m}/F_{60\mu m}$) (Fig.~\ref{seren})  
includes all galaxies with measurements or upper limits in the 60 and 
100~$\mu m$ IRAS bands .
The continous line has a slope of $\approx$~1 and corresponds to 
 the least square bisector regression for the 
ISOPHOT 170 $\mu m$ serendipity survey (ISS, \cite{sti00}). Most of the 
galaxies have a behavior that 
is consistent with those from the ISS, apart FN1-007 
(in the lower right corner of the  figure; this object is peculiar, see 
the discussion of SEDs later on).
Their colors  are well within the
sequence of the normal galaxy sample of \cite{dale01}, and correspond more 
to  the quiescent end of their classification
(i.e. the coldest galaxies). Note that for Arp220 and M82, 
log($F_{100\mu m}/F_{60\mu m}$) $\approx 0.03$,  and 
log($F_{170\mu m}/F_{100\mu m}$) $\approx 0.25$,  
  ratios that are  more typical of active galaxies.

\subsection{The 15/170 $\mu$m color }

Twenty four 
galaxies from our identified sample have both $F_{170\mu m}$ and $F_{15\mu m}$ measurements. 
We add to those   sources $FN1-013$ and $FN1-053$ which do not yet  have  a 
redshift measurement but have a clearly identified 15~$\mu$m counterpart.
From their $F_{170\mu m}/F_{15\mu m}$ ratio, the contribution of these 26 
galaxies  to the CIB can be estimated. \\
The observed color of the CIB can be bracketted from the work of 
\cite{elbaz02} and \cite{renau01} : 
$30 <CIB_{170\mu m/15\mu m} < 75$. 
On the other hand, the model of \cite{lagache04}, which reproduces 
(i) the number counts at 15, 24, 60, 70, 90, 160, 170 
 and 850~$\rm \mu$m, (ii) the known redshift distributions (mainly at 15, 
170 and 850~$\rm \mu$m),
(iii) the local luminosity functions at 60 and 850~$\rm \mu$m  and 
(iv) the CIB (from 100 to 1000~$\rm \mu$m) and its fluctuations
(at 60, 100 and 170~$\rm \mu$m), gives  a ratio $CIB_{170\mu m/15\mu m}$ = 
59.4. In view of the excellent
agreement of the \cite{lagache04} model with the observations from the 
mid-IR to the mm range, and the 
fact that the CIB color from the model lies well within  the range 
given by the 
observations, we   take this value from the model (60) as the best 
estimate of the 170/15 color of the CIB.\\

\begin{figure}
\begin{center}
\epsfxsize=9.cm
\epsfysize=7.cm
\epsfbox{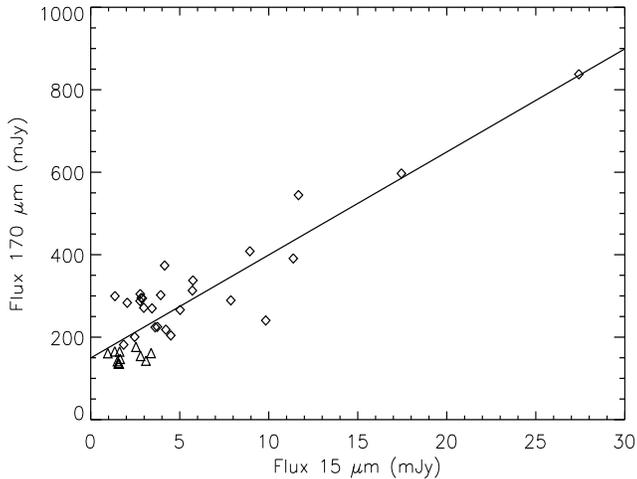}
\caption{\label{correl15_170} 170/15 correlation. The diamonds are for $S_{170}>4\sigma$
and the triangle for $3\sigma<S_{170}<4\sigma$. The continuous line
is the best fit for all sources. It has a slope of 170/15=25.}
\end{center}
\end{figure}

The 170/15 relation for the observed FIRBACK galaxies 
is shown in Fig~\ref{correl15_170}. For sources with $S_{170}>4\sigma$,
the slope of the correlation is equal to 22.3. If we add  to this sample
the sources with $3\sigma<S_{170}<4\sigma$, which are not identified 
optically, but
have clear 15~$\mu$m counterparts, the slope increases from 22.3 to 25, 
but is still more than a factor of  2 below the mean color of the CIB 
as discussed above. These sources  are thus clearly not representative
of the bulk of the sources contributing to the 170~$\mu$m CIB. 
This is not surprising, as (1) it was  shown by \cite{dole01} that 
 the bright FIRBACK sources (down to the  $3\sigma$ limit of the survey) 
 represent in practice less than 5$\%$ of the CIB at 170~$\mu$m and (2)
the observed CIB 170~$\mu$m/15~$\mu$m ratio is about three times higher than 
that of the galaxies studied here, and is  better  matched with higher 
z sources due to the K-correction (around z $\sim 0.8$, as observed by
\cite{elbaz02}). 

\subsection{Spectral Energy Distributions}

\subsubsection{Comparison with template spectra}
\begin{figure*}
\begin{center}
\epsfxsize=13.cm
\epsfbox{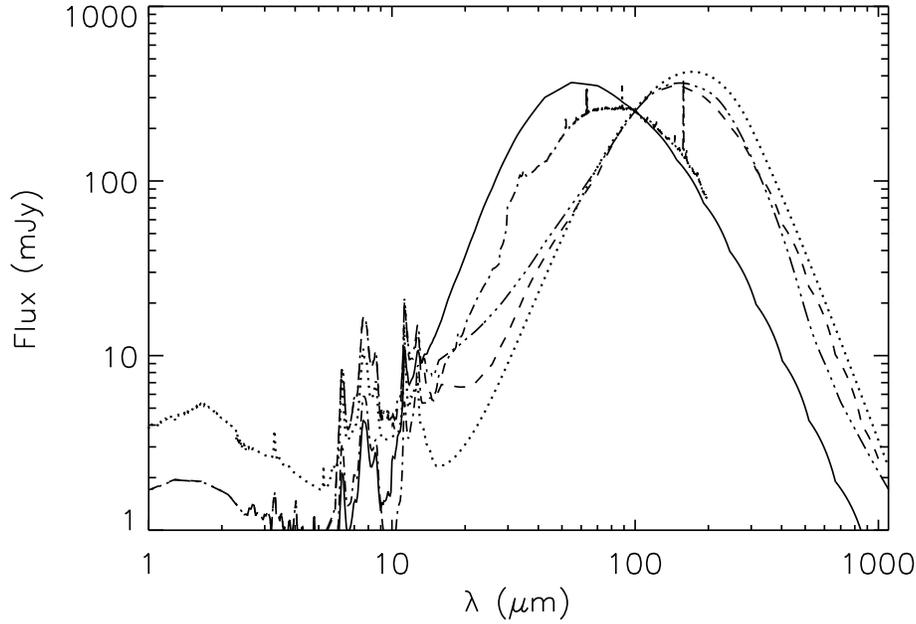}
\caption{Comparison of template  SEDs from Dale et al. (2001) ($\alpha$=2.6, 
quiescent case, dashed line; and $\alpha$=1, active case, continuous line)
and Lagache et al. (2003) ( ``cold galaxy'', triple-dot-long-dashed line) 
with  observed SED's. 
The dotted  line is the  observed SED of M101, the 
dashed-dotted  line the SED  of M82 ( curves  normalised at 100 $\mu$m). 
 \label{SedComp}  }
\end{center}
\end{figure*}

\begin{figure*}
\begin{center}
\epsfxsize=15.cm
\epsfbox{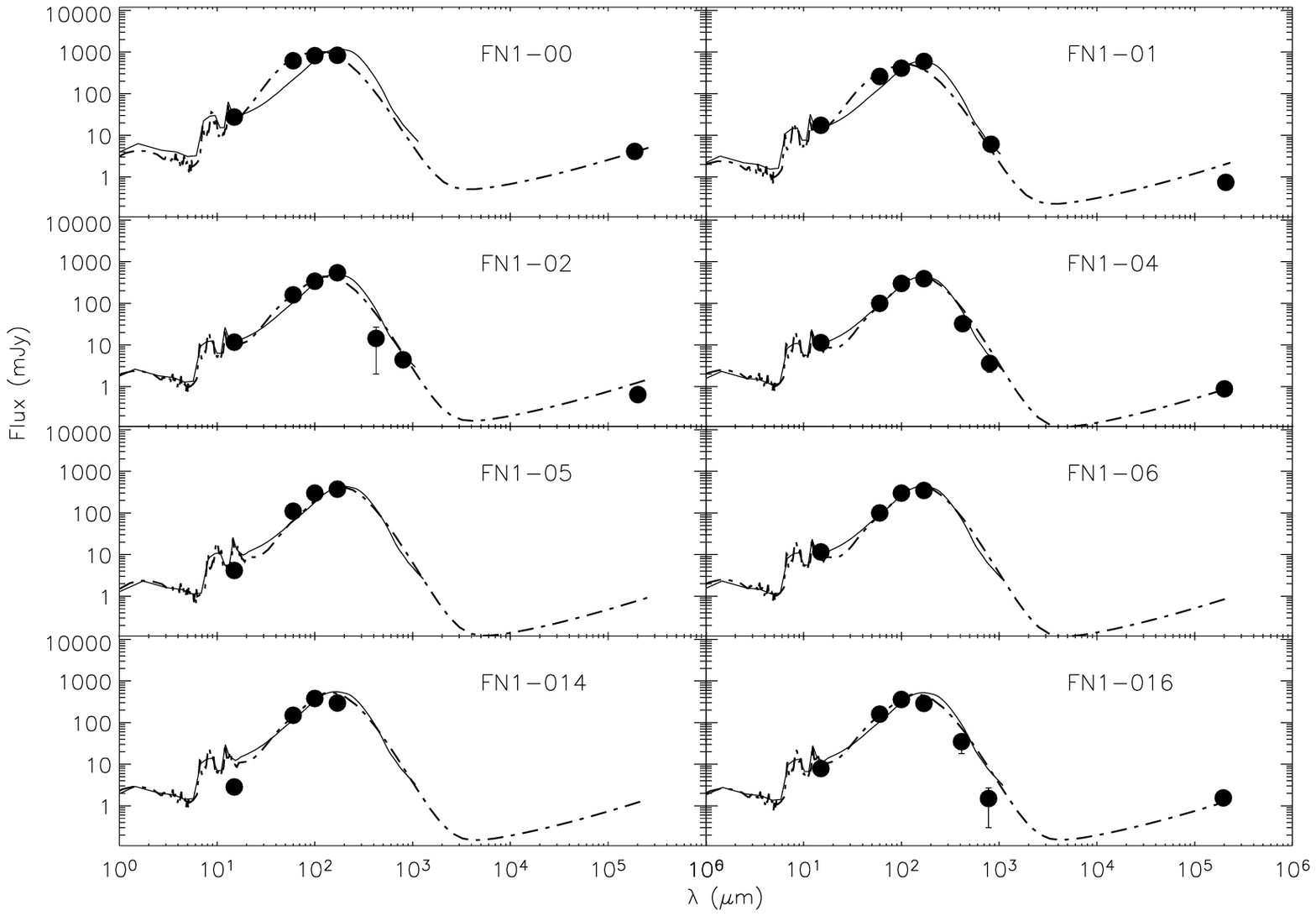}
\caption{\label{seddale}  SEDs  
  of those identified sources that have both measured  60~$\mu m$ 
and 100~$\mu m$ fluxes are fitted with a Dale model.  
The dashed-dotted line is the Dale model fit, 
the thin continuous one the ``cold galaxy'' spectrum template (Lagache et al. 
(2003)). 
Both the model and the template have been normalized at 170~$\mu m$. }
\end{center}
\end{figure*}

To derive an IR luminosity when only a few sampling points are available, the 
most reliable technique is to identify a template spectrum whose 
Spectral Energy Distribution (SED)  best matches the observations 
(although the solution might not be unique). There are plenty of SEDs that  
could be tried: the ones we used are shown in Fig.~\ref{SedComp} 
and are discussed below. 
Our first tests have shown (see also \cite{patris03}) 
that typical starbursts SEDs like (M82) did not match the data, and that colder 
components were required.  
\begin{figure*}
\begin{center}
\epsfxsize=15.cm
\epsfbox{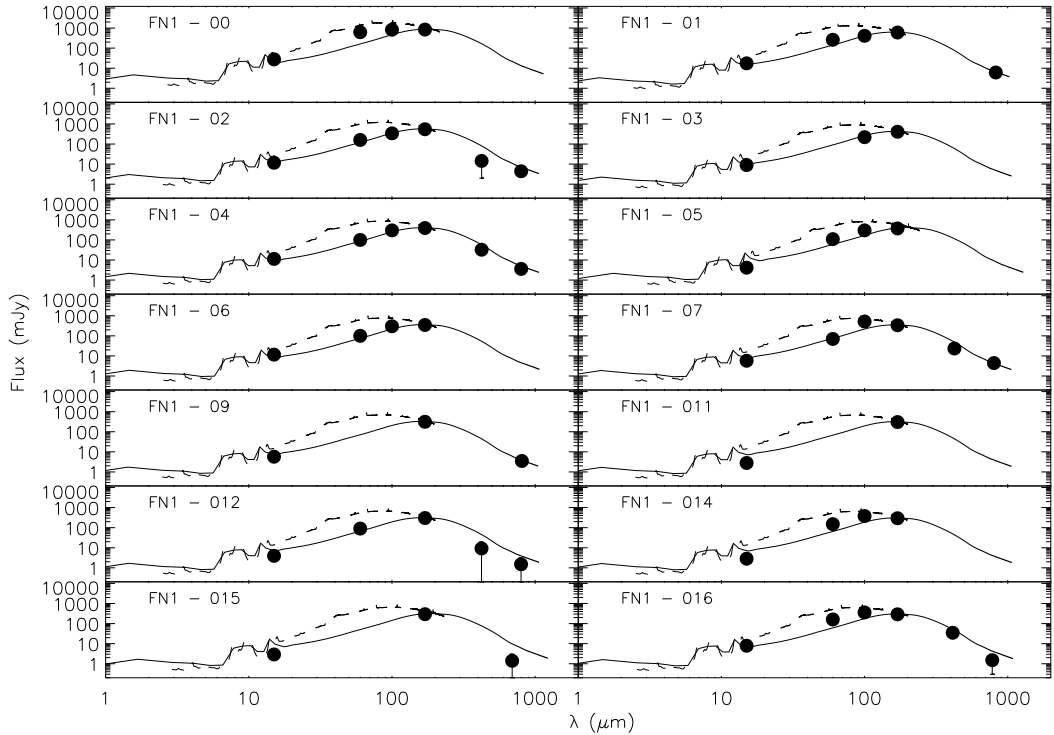}
\caption{The results from the fit of the source SEDs with a ``cold  
galaxy'' spectrum from Lagache et al. (2003). 
The continuous line is the ``cold galaxy'' spectrum template, the 
dashed line the spectrum of M82.     \label{sedcold1}  }
\end{center}
\end{figure*}
\begin{figure*}
\begin{center}
\epsfxsize=15.cm
\epsfbox{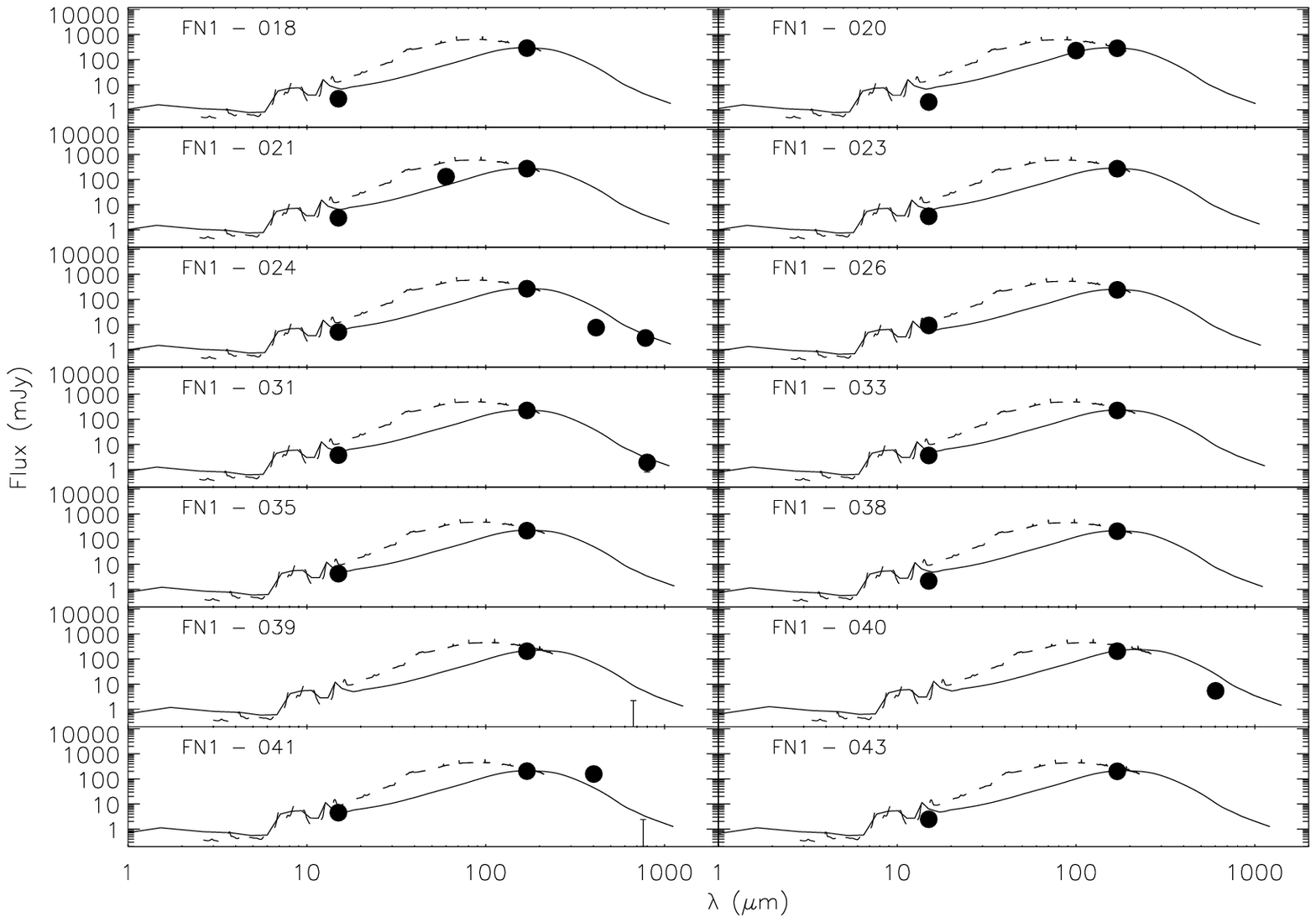}
\caption{  Same as in Fig~\ref{sedcold1} for the remaining identified 
sources.  \label{sedcold2} }
\end{center}
\end{figure*}
We thus used template spectra from Lagache et al. (2003), most 
notably their ``normal-galaxy'' spectrum, that is derived from observations 
of a sample of galaxies with measurements  
 from 15 to 850~$\rm \mu m$ (it was built  essentially 
from cold galaxies, including those from the present sample, and will 
therefore be called "the cold template" in what follows, to avoid confusion).   
This``cold-galaxy'' template has been fitted to each of our 
objects, and  is shown as  a solid line 
superposed on the observed points in Fig.~\ref{sedcold1} and Fig.~\ref{sedcold2}. 
The SED of the typical starburst galaxy M82  is 
also shown as a dashed line. 
The important point to note is that the M82 template, although fitting
quite well the 170/15 color, fails to reproduce the spectra at 60 and 
100~$\mu$m. This template, although widely used, is not appropriate
for the FIRBACK sources, which are colder galaxies. \\
For a quantitative appraisal of the contribution of various components, 
we use the models 
of Dale et al. (2001, 2002): they have modeled the infrared spectral energy 
distribution of 
normal star-forming galaxies, 
as a function of a parameter 
$\alpha$  which  quantifies the relative 
contribution of ``active'' and quiescent regions from galaxy to galaxy, 
$\alpha \approx 2.6$ being at the 
quiescent end and $\alpha \approx 1$ at the active 
end. 
Nine of our identified galaxies have both $F_{60\mu m}$ and 
$F_{100\mu m}$ measurements available 
and all but $FN1-007$ have a ($F_{60\mu m}/F_{100\mu m}$)  
ratio that falls in the range of colors modeled by Dale et al. (2001, 2002).
From the observed ($F_{60\mu m}/F_{100\mu m}$) ratio, we  derive  
 $\alpha$, and the  corresponding  best fit template spectrum.   
The best fit template spectra from Dale et al. (2002) are shown 
in Fig.~\ref{seddale}. 
The derived $\alpha$ (Table~\ref{tab-lum}) corresponds always to models where 
the quiescent component 
dominates ($\alpha > 1.6$): this clearly shows that our sample is composed of  
preferentially cold galaxies with only moderate star formation, 
rather than more active objects where the $F_{60\mu m}/F_{100\mu m}$ is 
usually greater than 1. \\
The chi squared of the two fits (the Dale et al. set or  the 
Lagache et al. ``cold galaxy'' spectrum; the M82 template is clearly not 
adequate, as can be seen from the figures) have been compared.
 The Dale et al. template spectrum is a better fit 
for only FN-000 and FN-016, while it  is comparable to the ``cold galaxy'' 
spectrum fit for the other 
galaxies. The FN-007 object remains an exception, but its optical spectrum 
does not give any indication of a peculiar nature. Its SED is however peculiar, 
with a very high  $F_{100\mu m}/F_{60\mu m}$ and a comparatively low 
$F_{170\mu m}/F_{100\mu m}$ ratio, difficult to explain with standard 
components. On average, the FIRBACK sources are clearly cold galaxies. \\ 

\subsubsection{IR luminosities}

Total  luminosities have then been calculated in several ways. First  by  
integrating the best fit spectra between 1 and 1000~$\mu m$. This was  done 
for all sources with the "cold" spectrum, even for those where only one measured 
point was available, namely the 170~$\mu m$ flux. The derived values are 
 shown in Table~\ref{tab-lum}. For those objects where a "Dale" spectrum could 
be determined, the luminosity is also derived (Table~\ref{tab-lum}).
Note that this second method was not applied to 
FN1-007, although it has measured  60~$\mu m$ and 100~$\mu m$ fluxes, because its 
 color lies outside  the range of validity of  Dale's models. 
Finally, we used also the formula proposed by Stickel et al. (2000) 
to  calculate luminosities when fluxes at 60, 100 and 170~$\mu m$ were 
available. 
Following their formula~:

\begin{equation}
F_{40-220} = 1.34 \times 10^{-14} \times \\
(2.58 F_{60 \mu m} + 
1.00 F_{100 \mu m} + 0.63 F_{170 \mu m} ) \quad \rm W m^{-2}
\end{equation}

\begin{equation}
F_{1-1000} = 1.35 F_{40-220} \quad \rm W m^{-2}
\end{equation}

\begin{equation}
L_{1-1000} = 4 \pi D^2 F_{1-1000}
\end{equation}

where D is the luminosity distance.
The luminosities derived with this formula are also shown in 
Table~\ref{tab-lum}. \\
For sources for which  60 and 100~$\mu m$ fluxes are available, 
the estimated luminosities can be compared.
 The luminosities estimated from  Dale's  best 
fit model are generally higher than the luminosities derived by the two other 
methods: this can be attributed to the fact that Dale's models  better take  
into account the contribution from the cold emission at very long wavelengths.  
But the differences are always smaller than a factor 
of two, and generally lower than 30 $\%$. In the subsequent analysis, 
we shall use (unless otherwise stated) the luminosity calculated from 
the Lagache model, as it is available for all objects. 


\begin{table*}[!htf]

\begin{flushleft}
\begin{tabular} {c|c|c|c|c|c|c|c|c|c} \hline 
ID& z& Lagache Lum.& Stickel L.&  Dale L. (and $\alpha$)& H$\alpha$ L.& K Lum.& SFR(IR)& SFR(H$\alpha$)& q\\ 
& & ($10^{10} L_{\odot}$)&( $10^{10} L_{\odot}$)& ( $10^{10} L_{\odot}$)& ( $10^{6} L_{\odot}$)&( $10^{9} L_{\odot}$) &$ M_{\odot}/yr$& $10^{-2}M_{\odot}/yr$& \\ \hline
     0  &  0.145  &  52.3   &73.0&  95.3 (1.69)&  222& 5.87    &  68& $>$675& 2.30\\ \hline
     1  &  0.030  &   1.5   & 1.3&   1.8 (1.83)&  4.9& 0.70    & 1.9&  $>$15& 2.69\\ \hline
     2  &  0.063  &   6.4   &4.6 &   6.1 (2.07)&  3.2& 2.32    & 8.3& $>$9.7& 2.60\\ \hline
     3  &  0.063  &   4.9   &    &             &  5.2& 2.54    & 6.4&  $>$16&     \\ \hline
     4  &  0.063  &   4.6   &3.4 &   5.2 (2.52)&  2.2& 4.97    & 6.0& $>$6.7& 2.32\\ \hline
     5  &  0.288  & 100.8   &70.5& 113.6 (2.35)&     &         & 131&       &     \\ \hline
     6  &  0.073  &   5.6   &4.4 &   7.2 (2.52)& 11.3& 2.80    & 7.3&  $>$34&     \\ \hline
     7  &  0.063  &   3.6   &3.8 &             & 12.8& 2.14    & 4.7&  $>$39& 2.35\\ \hline
     9  &  0.053  &   2.4   &    &             &  2.8& 3.14    & 3.1& $>$8.5&     \\ \hline
     11 &  0.063  &   3.6   &    &             &  3.0& 1.39    & 4.7& $>$9.1&     \\ \hline
     12 &  0.063  &   3.8   &    &             &  8.2& 0.93    & 4.9&  $>$25& 2.66\\ \hline
     14 &  0.07   &   4.1   &4.8 &   7.9 (2.25)&     &         & 5.3&       &     \\ \hline 
     15 &  0.23   &  48.6   &    &             &     & 4.23    &  63&       &     \\ \hline
     16 &  0.090  &   6.8   &8.5 &  13.3 (2.13)& 23.6& 3.02    & 8.8&  $>$72& 2.22\\ \hline
     18 &  0.067  &   6.0   &    &             &  3.3& 2.22    & 7.8&  $>$10&     \\ \hline
     20 &  0.01  &    0.1   &    &             &     &         &0.13&       &     \\ \hline
     21 &  0.064  &   3.2   &    &             &  3.6& 1.61    & 4.1&  $>$11&     \\ \hline
     23 &  0.063  &   3.1   &    &             &  5.9& 0.89    & 4.0&  $>$18&     \\ \hline
     24 &  0.086  &   5.7   &    &             &  8.5& 1.89    & 7.4&  $>$26&     \\ \hline
     26 &  0.077  &   4.3   &    &             &  1.7& 1.28    & 5.6& $>$5.2&     \\ \hline
     31 &  0.063  &   2.6   &    &             &  1.9& 1.15    & 3.4& $>$5.8&     \\ \hline
     33 &  0.133  &  12.4   &    &             &     &         &16.1&       &     \\ \hline
     35 &  0.137  &  11.9   &    &             & 32.5& 5.30    &15.5&  $>$99&     \\ \hline
     38 &  0.107  &   6.9   &    &             &  6.7&         & 9.0&  $>$20&     \\ \hline
     39 &  0.269  &  47.6   &    &             &     & 2.23    & 62 &       &     \\ \hline
  40$^a$&  0.45   & 122.9   &    &             &     & 0.17    & 160&   2000&     \\ \hline
     41 &  0.12   &   8.6   &    &             &     & 1.22    &11.1&       &     \\ \hline
     43 &  0.21   &  27.5   &    &             & 21.1& 7.23    &  36&  $>$64&     \\ \hline
     49 &  0.127  &   9.8   &    &             &     & 3.39    &12.7&       &     \\ \hline

\end{tabular}
$^a$: Data from  \cite{chapman02}, main component, reddening corrected H$\beta$ luminosity. 
\end{flushleft}

\caption{\label{tab-lum}
Source luminosities  calculated from the integrated ``normal-galaxy'' spectrum (Lagache Lum.), 
Stickel's formula (Stickel Lum.) and the best fit Dale spectrum (Dale Lum.) with 
its $\alpha$ value. The H$\alpha$ luminosities, uncorrected for 
extinction (hence the "$>$" sign in the corresponding SFR), are also given. 
The K luminosity is directly derived from the K magnitude. 
The "q" parameter is the logarithmic ratio of IR to radio flux, following 
\cite{helou85}. } 
\end{table*}

\section{The radio-FIR  relation}

The far-IR (FIR)-radio relation was introduced by \cite{helou85} from 
the study of normal galaxies with IRAS, and was shown to be quite general  
for various samples of galaxies (see \cite{condon92} for a review). The 
tightness of the relation, and its small dispersion, is best represented 
by the "q" parameter introduced by \cite{helou85}, which is the ratio of 
FIR (as measured by IRAS) to the radio continuum (1.4 GHz) flux densities: 

\begin{equation}
q = log [(FIR/3.75 \times 10^{12} \quad \rm W m^{-2})/(S_{\nu}/\rm W m^{-2} Hz^{-1})] 
\end{equation}

where FIR is the 40-120 $\mu$m flux density derived from IRAS measurements  
via: 

\begin{equation}
FIR/Wm^{-2} = 1.26 \times (2.58S_{60 \mu m} + S_{100 \mu m}) 
\end{equation}

The average value for q was found to be around 2.3 with a scatter lower than 
0.2 (\cite{helou85}). \\
We can calculate this  parameter  for those FIRBACK objects where IRAS data 
are available, and found similar values (Table \ref{tab-lum}): the objects 
 lie slightly above the mean relation,  with an 
average of 2.45 for 8 objects. 
 To increase the number of objects, one could calculate backwards 
 a FIR from the integrated IR luminosity (for instance from the "Lagache" 
luminosity which is available for all of them, see previous section):  
this would then represent a $FIR_{1-1000}$ 
instead of $FIR_{40-120}$ and, as expected, the q ratio is increased 
accordingly.  
But  only few  objects are added this way (due to the 
limited radio detections), so that  no advantage is obtained  in 
practice from this different approach.  \\ 
The main result  is that no object has 
a q value significantly lower than the average 2.3 found in normal galaxies: 
there is therefore no evidence for a significant contribution to the total 
energy balance from a radio-loud AGN in any of those objects. For those few 
objects where the presence of an AGN is indicated from the spectroscopic data, 
and where radio data are also available, the q ratio is not higher, 
suggesting that the contribution of the AGN to the FIR flux is also 
negligible. However, we  cannot exclude a contribution 
from a radio-quite AGN.  \\ 

\section{Star formation rates}

Several indicators can be used to estimate the Star Formation Rate (SFR) of 
galaxies, but it is widely accepted that the far-IR luminosity gives the best 
estimate, because it measures the bolometric luminosity of the object: 
this is correct  as long 
as one can be sure that a hidden AGN is not significantly contributing to 
the total energy output. 
We have used the IR luminosity computed from the fit of the "Lagache" template
(which is available for all the identified objects) and derived the SFR with 
the calibration of \cite{devr99}: \\  
$$ \rm{SFR\:(M}_\odot\rm{per\:year)} = 
\frac{\rm{L}_{IR}\:(10^{9}\rm{L_\odot})}{7.7}$$  
 
The result is given in Table~\ref{tab-lum}. 
Most of the objects have star formation  rates of only a few $M_{\odot}$/year 
and are therefore moderate starbursters only. Five objects  are 
however standing out: $FN1-000$, $FN1-005$, $FN1-015$, $FN1-039$, and $FN1-040$. 
$FN1-005$ and $FN1-040$ are the most 
distant objects in this limited sample of fully identified sources, and are 
clearly  ULIRG's detected at larger distance than the average, bright, FIRBACK 
sources (for details on $FN1-040$, see \cite{chapman02}). The same is true 
for $FN1-015$ although it has a three times smaller 
IR luminosity. For $FN1-039$ we have no detailed information.  More 
interesting is the case of $FN1-000$, which is  
closer, and at similar distances than, for instance, $FN1-035$ or $FN1-033$, 
but has a five times greater luminosity. We have no evidence  
that an AGN is located inside this galaxy neither from spectroscopy nor from 
the q parameter, so we consider it as a vigorous, 
local starburst galaxy. \\
 We can, in principle, also use the H$\alpha$ luminosity to derive the SFR, 
following the recipe of, e.g., \cite{kenn92}: \\
$$ \rm{SFR\:(M}_\odot\rm{per\:year)} = 
\frac{\rm{L}_{H\alpha}\:(10^{41}\rm{erg/s})}{1.26}$$ \\ 
 The result is also given in Table~\ref{tab-lum}. 
 For the 17 objects where we have both IR and 
H$\alpha$ estimates, the latter always gives  a much lower SFR, on  average  
 a factor of 22 lower than the former one (note that the 4 high-luminosity 
objects are not included in this sample). 
This is clearly due to the fact that our 
H$\alpha$ fluxes are not  corrected for internal extinction. 
We have no reddening estimate for each individual galaxy.  
In a few cases only, like $FN1-000$, an extinction can be estimated 
from the H$\alpha$ over H$\beta$ ratio (where we find  $A_{v}=2$, with 
the necessary caveat due to the unresolved underlying absorption lines). 
We could apply to  
the whole sample  an  average extinction correction of $A_{v}\sim~1$, 
as was done in other 
 cases (e.g.  \cite{pett01} where a low  extinction was estimated from the 
comparison between UV and visible properties or \cite{lilly03} where this 
value was derived from 
a local reference sample). It is however very unlikely that such a low 
extinction would be appropriate for our sample which is mainly emitting in 
the far IR. Instead, we can use a value 
obtained from a very 
similar sample where more data are available, namely the FIRBACK southern 
field (\cite{patris03}): the average extinction found there is  $A_{v}=3$.  
This  
corresponds to an upwards correction of a factor of 10 to the SFR derived 
from the H$\alpha$ luminosities. While such a correction is only indicative, 
because the dispersion of extinction values from object to object may be very 
high, the  
similarity of these two samples ensures  that the selection effects 
are minimized: this value is certainly more appropriate than the standard  
$A_{v}=1$ value applied in other samples. After such correction, it appears 
 that  the SFR derived from H$\alpha$ becomes closer to the one derived 
from the IR, although still systematically lower. The remaining difference 
is  due in part to some slit losses and, more probably, to a large optical 
thickness in H$\alpha$ thatprevents an accurate  extinction correction. 
 This problem is similar to the one encountered 
in other IR selected samples (like in \cite{patris03}) but without additionnal 
 data, the comparison between the various SFR estimators cannot be pursued
further. \\ 
The radio-FIR relation, discussed in the previous section, shows however 
that the "q" parameter for  
these galaxies (at least the 8 for which it could be determined) is not 
significantly different from  the 
standard value found by \cite{condon91} in the Bright Galaxy Sample 
of \cite{soifer89}: 2.4 for our objects instead of 2.3 in the BGS. Our 
galaxies, with the exception of the 5 cases mentioned above, 
 are thus effectively moderate starbursters only. 
 
\section{Discussion and conclusions}

Out of the 56 FIRBACK N1 sources with 170 $\rm \mu$m fluxes above the 
4$\sigma$ limit of 180 mJy, 28 have been firmly identified and 17 others 
have at least one of the contributors identified. Only 11 sources remain 
unclear, essentially due to the lack of additional data (radio or near-IR) 
with better positional accuracy.  
Most of the identified sources are  quiescent star-forming galaxies 
which  exhibit a colder spectrum than a standard  IRAS starburst galaxy. 
There is  no overlap  with the catalogue of galaxies 
resulting from the ISOPHOT 170 $\rm \mu$m Serendipity Survey (\cite{sti04}), 
although in view of their detection limit of $\sim~0.5 Jy$ one would have 
expected that the first 3  objects from the FIRBACK-N1 catalogue could have
 been found there. We can nevertheless compare the properties of our objects 
that have also detected IRAS fluxes with those from the Serendipity Survey 
discussed by \cite{sti00}. 
Both samples have similar far-IR colours, with an $F_{100 \mu m}/F_{60 \mu m}$ 
ratio greater than $\sim 1.5$ and a $F_{170 \mu m}/F_{100 \mu m}$ greater than 
0.7, which we qualify as "cold" galaxies, dominated by a cold dust component 
with temperatures roughly between 20 and 40K. This is clearly different from 
the "warm" LIRGs and ULIRGs, whose $F_{100 \mu m}/F_{60 \mu m}$  ratio is 
generally close to or smaller than 1, and which are strong starbursters. 
The bright FIRBACK galaxies seem therefore to represent a fainter version of 
the "cold" galaxies detected in the Serendipity Survey. \\
In terms of optical spectral properties, they resemble the faint IRAS 
galaxies selected  by \cite{bertin97} and described in \cite{patris03}: 
emission lines, moderate to strong reddening and moderate SFR. There is 
unfortunately no IR colour information available for the  \cite{bertin97} 
sample, for comparison, 
as those galaxies represent the faintest objects detected in the IRAS survey, 
in the most sensitive channel ($F_{60 \mu m}$ between 150 and 250 mJy), and 
are  thus not detected in the other channels. 
Its seems however clear, from the three 
different samples just discussed, that a large population of "cold" galaxies 
exists at least in the local universe (with a few tens of objects per square 
degree),
 with moderate star formation, whose 
contribution to the global star formation rate is probably significant, 
although it has  been rather neglected until now, in the "rush" to find always 
more extreme starburst objects. \\

The bright FIRBACK galaxies studied here are mostly nearby, Luminous IR 
Galaxies (LIRGs) 
with moderate star-formation rates (about 5 to 10 M$_{\odot}$/yr) and 
are not particularly associated with detectable merging or interacting
systems (only three mergers are clearly detected  among the 28 fully 
identified objects). 
One could question whether they should be called "starburts galaxies" 
or whether they do not simply represent disk galaxies with more quiescent 
star formation (or disks ionised by diffuse radiation). 
Recent Spitzer observations of classical spiral galaxies like 
the Scd spiral NGC 300 (\cite{helou04}) or 
or the Sc M33 (\cite{hinz04}) show that their 
integrated SFR is low, of the order of 0.1-0.2M$_{\odot}$/yr, and that, 
although their 160$\mu$m emission might be more diffuse, 
the shorter wavelength emission 
is clearly associated with ionising stars. Such SFR are much smaller than the 
ones obtained for our FIRBACK galaxies. 
A larger sample of galaxies has 
been observed in H$\alpha$ by \cite{james04} to derive SFRs: they find an 
integrated, extinction-corrected  SFR of 1-3 M$_{\odot}$/yr 
for the most active spirals, the types Sbc or Scs. 
Our H$\alpha$ rates are difficult to compare directly because of the 
uncertain extinction correction: but either applying the average correction 
of 10 discussed earlier, or using individually corrected H$\alpha$ rates from 
the similar, southern sample discussed by 
\cite{patris03}, or using the  FIR-SFR which correlate well with the corrected  
H$\alpha$ rates as shown by \cite{patris03},
we have here SFRs on average 3 to 4 times greater than the most 
active galaxies of \cite{james04}, the Scs. Furthermore, our morphologies are 
generally of earlier types than Sc (with often a central, bulky component),
 types for which the SFRs measured in  
\cite{james04} are comparatively smaller. 
By comparison, M101, a nearly face-on spiral whose SED, shown in 
Fig.~\ref{SedComp}, is  comparable to the quiescent one of 
\cite{dale01}, has a FIR-SFR  of 5.7M$_{\odot}$/yr   
(for an IR luminosity of 4.4 10$^{10}$ L$_{\odot}$, derived from IRAS data), 
similar to the ones of our FIRBACK galaxies. But its classification  
is SBc, again of later type  
than most of our galaxies, and furthermore with a bar, 
usually believed to be linked  with larger SFRs than in non-barred galaxies. 

The IR luminosity function derived from the Bright Galaxy Sample by 
\cite{soifer87} shows that the break occurs at a luminosity of 1.7 10$^{10}$ 
L$_{\odot}$. Following \cite{sti00}, this value, calculated over the 
40-220 $\mu$m range, converts into roughly 3 10$^{10}$ L$_{\odot}$ for the 
3-1000 $\mu$m range discussed in this paper: the majority of our objects are 
therefore typical L$^{*}$ objects, or slightly above. Noticable exceptions 
are FN1-0, 5, 15, 39, 40  or 43, which are much more luminous. 
When comparing with the local K-band luminosity function of \cite{cole01}, 
most of our galaxies are  only slightly brighter than the M$^{*}_{K}$ 
magnitude of -24.2 (with H$_{o}$ = 70 km/s/Mpc as used here) 
(the exceptions being FN1-1, 12, 23 and 40 which are significantly fainter), 
so that they 
are  not extremely massive  galaxies. The ratio of total far-IR 
luminosity to the K-band luminosity, which can be interpreted as an indicator 
of the ratio of present to past star formation, is rather homogeneous 
for the sample, with values between 20 and 40, showing that the present SF  
dominates the energy output. 
A few  objects are more active than average (FN1-1, 23, 39, 40, 41), but are 
also less massive (with the exception of FN1-39).  The outstanding object is  
   FN1-40 (which is also the most distant object of the 
sample), appearing as  a sub-L$^{*}$ object from its K magnitude, 
but heavily forming stars, and with a far-IR luminosity bringing it into 
the ULIRG class: this galaxy  is 
therefore clearly of a different nature than the average bright FIRBACK 
galaxy and is probably a dwarf galaxy experiencing  
one of its first starbursts; see also  \cite{chapman02}. 
 We conclude that, while a few 
 are standard,  disk-dominated spirals (like for instance FN1-02),  
many of these objects    seem   to have a larger SFR  than standard 
spirals, with a concentration towards the central regions which could indicate 
the final phase of a former merger event. 

As far as the CIB is concerned, it is  clear  that  the bright FIRBACK 
galaxies   have mid- to far-IR colors clearly different from the mean colors 
of the CIB.  
They do therefore   
  not represent the bulk of the sources contributing 
to the mid- and far-IR CIB,  
contrary to earlier expectations (e.g. \cite{puget99}, or \cite{devr00}).
The main contributors to the IR-CIB are thus expected to be more 
distant, more active  galaxies, which will  possibly be detected within the 
fainter part of the FIRBACK survey.  \\

This local cold population  however has to be taken into account in the 
modelling of the evolution of IR galaxies. 
In practice, it  strongly affects the redshift
distribution of  170 $\rm \mu$m sources predicted by the models. 
When models consider  only starburst galaxies, 
they lead to a redshift distribution for the 4$\sigma$ 
FIRBACK galaxies that is clearly biased towards redshifts much  higher 
than those observed here (e.g. \cite{devr00}). 
Our present results show that the number counts 
 are dominated  by local cold galaxies
for 170 $\rm \mu$m fluxes greater than about 240 mJy. 
This is now well taken into account  in  the   Lagache et al. (2003)
 models,  which  
predict an equal contribution of starburst- and of ''cold'' galaxies 
at 250 mJy. This is also in agreement with the statistical results derived 
from sub-mm observations by \cite{sajina03}. \\

 The detailed analysis of  the FIRBACK population  obviously needs 
 to be completed by the identification  of the fainter, more 
distant counterparts. This will require  
 many complementary observations in various wavelength ranges, from the 
ground with larger facilities, and from space. But, 
as the confusion  limit of the FIRBACK survey clearly restricts the 
detection to relatively nearby objects,  results from   {\it SPITZER}
should notably improve the situation.  
Indeed, the first number counts at 24 $\rm \mu$m  in the N1 field by  
e.g. \cite{chary04} suggest a strong contribution from luminous IR galaxies 
in the redshift range between  0.5 and 2.5. 
Inclusion of the shorter wavelengths data from IRAC will also allow to better 
determine the contribution of the quiet, diffuse component 
to the overall energy output in the nearby objects.   
A detailed study of those 
galaxies (which is underway; e.g. Sajina et al., in preparation) 
is therefore likely to bring new light on the evolution of IR 
galaxies and their relation to the higher-z sources found in sub-mm 
observations. \\

Acknowledgements: We  thank  Jean-Loup Puget for enlightening 
discussions and support during this work;  P. Chanial for providing 
the M101 data in advance of publication;  S.C. Chapman  
 for providing us two  redshifts (FN1-5, 39) and  D. Dale for providing his  
SED models.  
S.M. acknowledges  support from the ESA External Fellowship program, and advice 
from Marc Postman for the morphological classification of the galaxies.  \\
This paper is based on observations made with ISO, an ESA project with 
instruments funded by ESA Member States and with the participation of 
ISAS and NASA. \\
This publication makes use of data products from the Two Micron All Sky Survey, which is 
a joint project of the University of Massachusetts and the Infrared Processing and Analysis 
Center/California Institute of Technology, funded by NASA and NSF. It also 
used data from the Lyon Extragalactic Database (LEDA).

\begin{figure*} 
\includegraphics[width=6cm,angle=270]{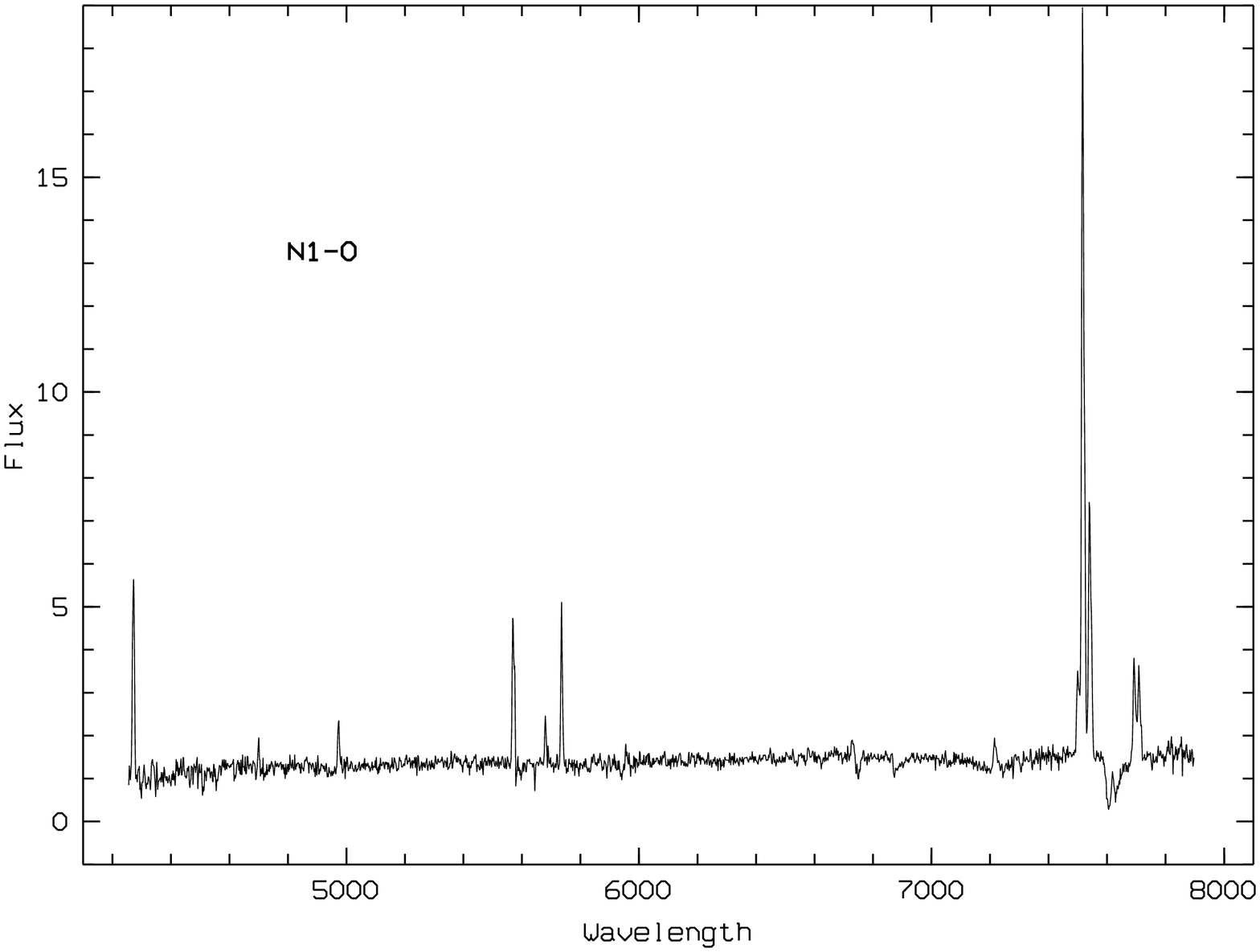}
\includegraphics[width=6cm,angle=270]{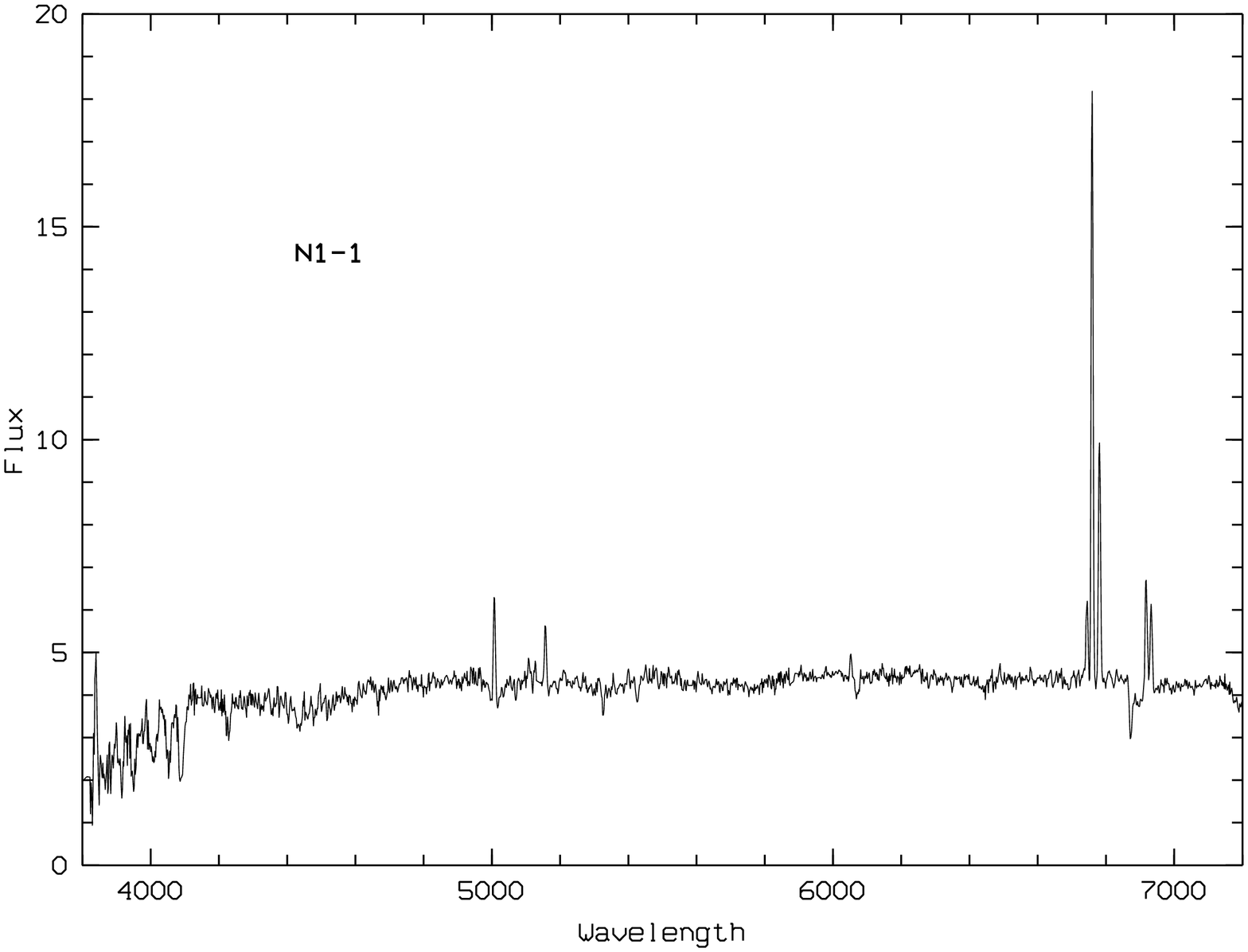}
\caption{ Spectra of individual objects as examples (FN1-0 and FN1-1).  
Spectra of further  objects are only available in electronic form. }
\end{figure*}

\end{document}